# Decoding Energy Decomposition Analysis: Machine-Learned Insights on the Impact of the Density Functional on the Bonding Analysis

Toni Oestereich, Ralf Tonner-Zech*, Julia Westermayr*

Wilhelm-Ostwald-Institut für Physikalische und Theoretische Chemie, Universität Leipzig, 04103 Leipzig, Germany

corresponding authors: Prof. Dr. Julia Westermayr: Julia.westermayr@uni-leipzig.de; Prof. Dr. Ralf Tonner-Zech ralf.tonner@uni-leipzig.de

The concept of chemical bonding is a crucial aspect of chemistry that aids in understanding the complexity and reactivity of molecules and materials. However, the interpretation of chemical bonds can be hindered by the choice of the theoretical approach and the specific method utilized. This study aims to investigate the effect of choosing different density functionals on the interpretation of bonding achieved through energy decomposition analysis (EDA). To achieve this goal, a data set was created, representing four bonding groups and various combinations of functionals and dispersion correction schemes. The calculations showed significant variation among the different functionals for the EDA terms, with the dispersion correction terms exhibiting the highest variability. More information was extracted by using unsupervised learning in combination with dimensionality reduction on the data set. Results indicate that, despite the differences in the EDA terms obtained from different functionals, the functional has the least significant impact, suggesting minimal influence on the bonding interpretation.

# 1 Introduction

Chemical bonding is a fundamental concept in chemistry that plays a critical role in understanding the complexity of molecules and materials. Despite its significance, there is no unique definition of "chemical bonding" or what is a chemical bond. It thus remains a constructively but often controversially discussed topic in the chemical community.[2,3] Energy Decomposition Analysis (EDA) is one way to provide a more intuitive understanding of chemical bonding of molecules based on quantitative analysis.[4-8] EDA can be thought of as dissecting a puzzle to understand how each piece contributes to the overall structure, which means it comprises a decomposition of the total energy into its contributing components. A better understanding of the different contributions to a chemical bond can provide insights into what governs stability and reactivity.[7] Subsequently, this information can be used to improve the design of new molecules and materials, for instance, by manipulating the molecular structure such that specific chemical interactions can be altered in a desired way, e.g., by reducing Pauli repulsion or increasing orbital interaction.[1,9,10]

Although density functional theory (DFT) based approaches are widely used for EDA, the most used approximations have well-known limitations. One such approximation in the Kohn-Sham approach to DFT is the choice of the density functional, which is used to approximate the exchange-correlation energy, a crucial component of the total energy in DFT. Due to the unknown exact functional form, a large range of functionals have been developed over the last decades, resulting in different quantitative results for the same system.[11,12] One of the most commonly used type of density functionals in EDA studies are based on the Generalized Gradient Approximation (GGA). Particularly, the Perdew-Burke-Ernzerhof (PBE)[13] functional has often been employed. The PBE functional, combined with a method to account for dispersion interactions, has been shown to provide good results for a wide range of systems and has become a standard choice for many applications in theoretical chemistry, especially in the surface and material sciences.[14,15] However, for certain systems, other functionals may provide more accurate results, and the choice of the functional should be carefully considered for each specific system being studied.[16] As a result, the use of different functionals across various EDA studies could result in bonding interpretation becoming dependent on the specific choice of the functional. However, it is critical that the bonding analysis is consistent. Further, one wants to make sure that the bonding interpretation for a given system does not depend on the computational setup, most importantly the density functional, used to conduct the EDA analysis. In a previous study, we found the relative importance of dispersion attraction and covalent bonding contributions to depend on the density functional used.[17]

In this study, we systematically shed light on the impact of the choice of the functional on the results of chemical bonding analyses with EDA. We thus carried out EDA studies of 11 molecules that are representatives of the different types of chemical bonds in molecular systems: Covalent bonds, main group donor-acceptor bonds, transition metal donor-acceptor bonds, and hydrogen bonds. We used different density functionals, i.e., 7 GGA, 6 meta-GGA, 2 hybrid, 3 meta-hybrid, and 1 range-separated functional with various types of dispersion corrections resulting in 28 different calculations per molecule. The results are analyzed with the help of dimensionality reduction and supervised learning techniques. The different functionals are accounted for by encoding them into the molecular descriptor of the learning models. Analysis of the importance of the contributions of the descriptor provides insights into what constitutes the chemical bond and what impact the functional has on the result of the bonding analysis.

# 2 Theory

## 2.1 Energy decomposition analysis (EDA)

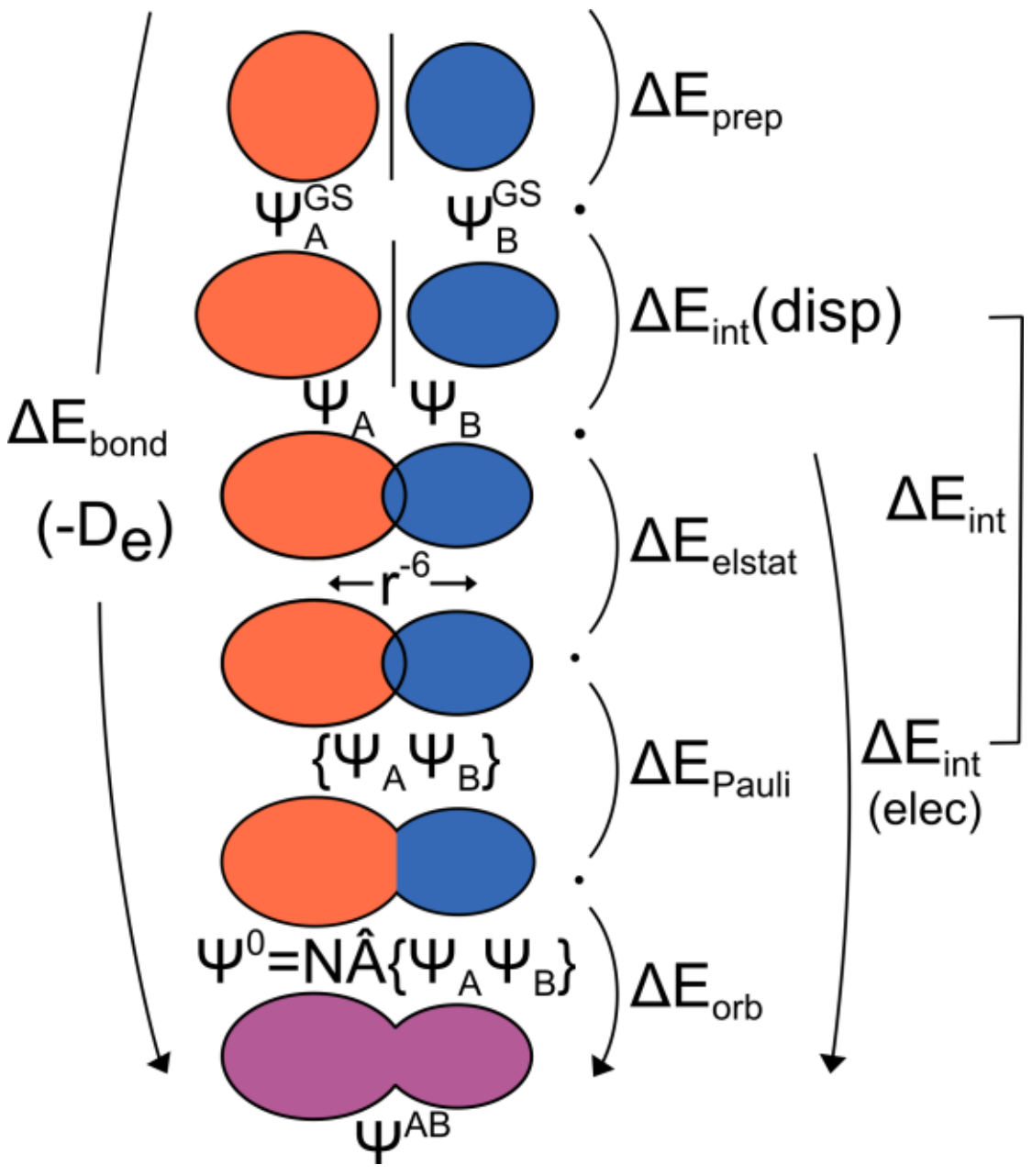

**Figure 1.** Schematic representation of energy terms derived in the EDA approach used to analyse a chemical bond between two generic fragments A and B forming AB. Adapted and modified based on reference 1.

The EDA procedure starts by defining molecular fragments, A and B, of a chemical system, AB, with an interaction energy, $\Delta E_{int}$. This interaction energy can be obtained by subtraction of the preparation energy, $\Delta E_{prep}$, from the bonding energy between the two fragments, $E_{bond}$ (eq. (1)).

$$E_{bond} = \Delta E_{int} + \Delta E_{prep} \qquad \text{eq (1)}$$

Next, $\Delta E_{int}$ is decomposed into dispersion, $\Delta E_{int}$(disp), and electronic, $\Delta E_{int}$(elec), terms (eq. (2)). The first term, $\Delta E_{int}$(disp), gives the fraction of dispersion energy contribution to the bonding which is usually taken from a semiempirical correction scheme. EDA then decomposes $\Delta E_{int}$(elec) into three well-defined terms that can be interpreted in chemically meaningful ways (eq. (3)). These terms are (1) the quasiclassical electrostatic interaction energy between the charge densities of the fragments, $\Delta E_{elstat}$, (2) the exchange repulsion between the fragments due to Pauli's principle, $\Delta E_{Pauli}$, and (3) the energy gain due to orbital mixing of the fragments, $\Delta E_{orb}$. In this way, information about the importance of electrostatic, dispersion, repulsion, and orbital interactions can be obtained.[18]

$$\Delta E_{int} = \Delta E_{int}(disp) + \Delta E_{int}(elec) \qquad \text{eq (2)}$$

$$\Delta E_{int}(elec) = \Delta E_{elstat} + \Delta E_{Pauli} + \Delta E_{orb} \qquad \text{eq (3)}$$

Although EDA based on Kohn-Sham DFT is the most widely used approach for bonding analysis, other EDA implementations exist. Some examples are the block-localized wavefunction (BLW) EDA,[19] the very similar absolutely-localized molecular orbital (ALMO) EDA[20-22] or the generalized Kohn-Sham (GKS) EDA.[23] A more comprehensive overview can be found in recent review articles.[7,9,18,24,25] Although few of these approaches have been extended to wavefunction based methods,[23,26] the application of DFT approaches by far prevails. The two main reasons are (i) efficient computations are possible with DFT approaches and (ii) the ease of interpretation resulting from the mean-field approach resulting in an effective one-electron picture.

## 2.2 Machine learning (ML)

ML comprises a collection of algorithms that aim to map an input space $X$ to an output target space $Y$ using statistical methods, expressed as $f: X \rightarrow Y$. Unlike traditional physical models, ML algorithms usually do not rely on prior physical assumptions but find functional relations by training on a set of reference data, containing both, $X$ and $Y$, in case of supervised learning, or only $X$ in case of unsupervised learning. Once trained, ML is powerful as it can generalize to new, unseen data, with high computational efficiency and the accuracy of the reference method.[27,28] In this work, we made use of the propensity of ML to find relations between molecular inputs, functionals, and EDA terms to analyze their influence on the bonding analysis. In the following, we will lay the groundwork for the terminology and concepts used in this work and provide a brief overview of the different types of ML algorithms employed.

### 2.2.1 Molecular representation

A critical component of ML is data representation, which involves encoding molecular structure and properties into numerical descriptors that can be used as input features, $X$, for ML algorithms. There are several types of descriptors available, including molecular fingerprints, SMILES strings, or descriptors based on 3-dimensional arrangements of atoms in a molecule.[29,30]

Since EDA focuses on bonds, local descriptors to represent the two atoms participating in the bond under investigation were used. Local descriptors can describe molecules of arbitrary sizes by representing atoms in their chemical and structural environment.[31] In particular, the Smooth Overlap of Atomic Positions (SOAP)[32] descriptor was employed here. SOAP transforms the atomic positions into a three-dimensional grid and computes the overlap between neighboring atomic densities. This procedure gives a set of coefficients that can be used as input features for an ML model, which remain unaffected by symmetry operations performed on atoms in the environment. In addition, we tested a minimal descriptor based on distances of the bond under investigation and elemental charges of the atoms involved in the bond as the bond distance separates the different bonding groups (see Figure S1 in section S1 in the supporting information (SI)).

Besides structural representations, also molecular properties can be used as input information to an ML model.[27] Therefore, the EDA terms and information on density functionals served as inputs.

### 2.2.2 Supervised learning

In this work, classification algorithms were used as supervised learning techniques, for which the model required access to the target properties, $Y$, which are EDA terms (regression) and bonding classes (classification) in this case.

For classification, we tested the different methods mentioned above that were used for regression to classify molecules into their bonding characters by using EDA terms, information on the method, and structural features as inputs, $X$.

### 2.2.3 Unsupervised learning

To unravel relations in input data, unsupervised learning was used. In contrast to supervised learning, unsupervised learning involves learning from unlabeled data, hence the model does not know about $Y$. This type of learning can be used for tasks such as clustering and dimensionality reduction.

Here, dimensionality reduction, i.e., principal component analysis (PCA), was used in this work to reduce the dimensionality of structural descriptors (used as $X$) of the chemical systems under investigation as these often contain many thousands of values. PCA is a linear method that finds the most significant orthogonal directions of variation in the data and projects the data onto a lower-dimensional subspace. The new set of variables are called principal components (PCs) and capture the variance in the original data. Noteworthy, PCs do not have a physical meaning, although feature importance analysis can be used to assess the influence of each descriptor value on a given PC.[33]

In addition, clustering tools were tested to assess the propensity to group data based on their structures and EDA terms without having access to the predefined bonding classes. Clustering can thus be seen as the unsupervised counterpart to classification.[29]

# 3 Models and Methods

## 3.1 EDA data set: Choice of molecules

To study the impact of the functional and dispersion correction on EDA results, a data set was generated that comprises 11 molecules (see Figure 2) and EDA terms obtained from 28 different methods for which the functional was varied (details on the different combinations of functionals and long-range interactions can be found in Table S1 in the SI). In total, 308 data points with each data point containing 8 entries obtained from EDA were generated. The molecules used are representatives of four characteristic bonding types, i.e., covalent (**1-3**), main group donor-acceptor (**4-6**), transition metal donor-acceptor (**7-9**) as well as hydrogen-bonded systems (**10**, **11**). The choice of these four classes was made to represent the most common bonding types. Covalent bonds result from both fragments contributing one electron each to the bond (shared-electron bonding). Donor-acceptor bonds result from both bonding electrons stemming from one fragment, only. The donor-acceptor bonds are further split into two groups, i.e., those of main group elements and those of transition metals. Although the bonding mechanism is the same, the elements and orbitals involved in the bonds are chemically different enough to distinguish them via EDA. Non-covalent interactions are exemplified by hydrogen-bonded systems.

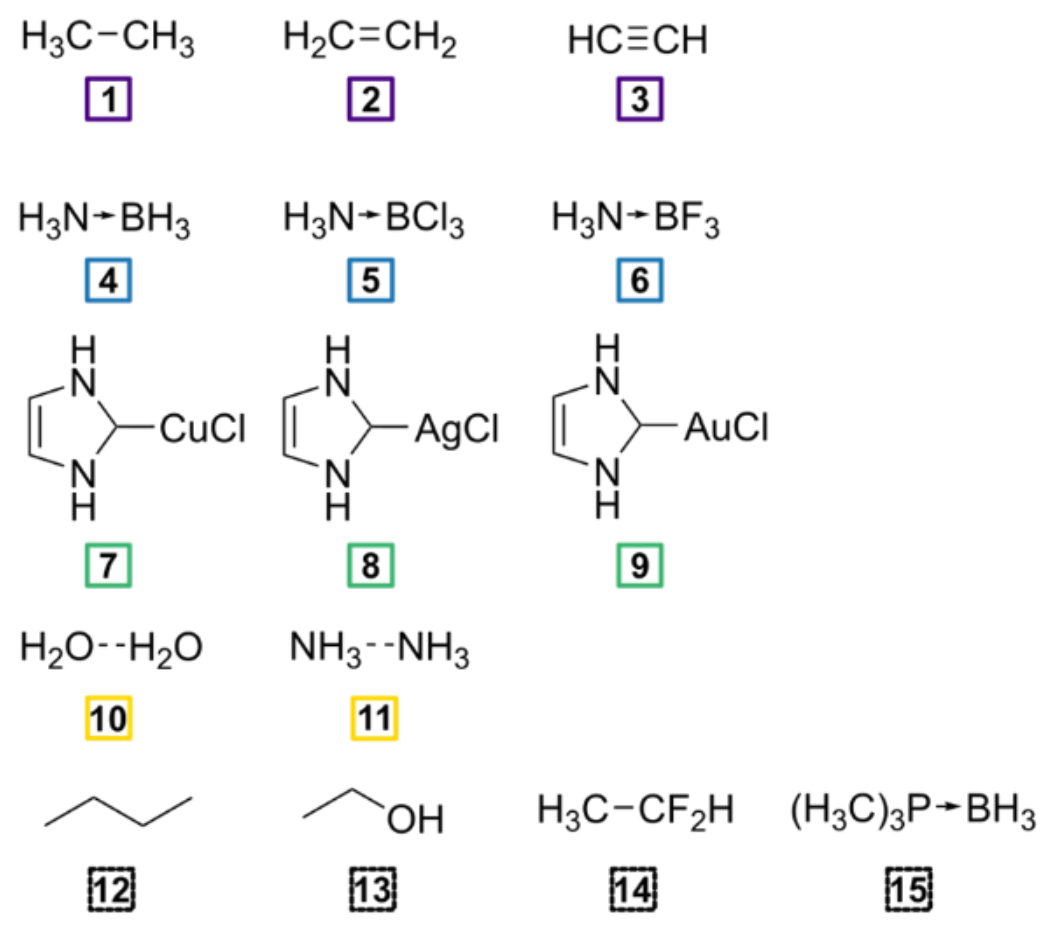

**Figure 2.** The molecules (**1-15**) investigated in this study. Molecules **1-11** comprise the training set. They represent the bonding classes of covalent bonds (**1-3**), main group donor-acceptor bonds (**4-6**), transition metal donor-acceptor bonds (**7-9**) and hydrogen bonded systems (**10**, **11**). Molecules **12-15** are included for testing only.

The molecules tested (Figure 2) are ethane (**1**), ethene (**2**), ethyne (**3**), ammonia borane (**4**), ammonia boron trichloride (**5**), ammonia boron trifluoride (**6**), imidazol-2-ylidene (nNHC) transition metal complexes of copper-, silver-, gold-chloride (**7-9**), as well as a water (**10**) and ammonia dimer (**11**). Subsequently, we generated a test set of additional molecules comprising butane (**12**), difluoroethane (**13**), ethanol (**14**), and trimethylphosphine borane (**15**) for which EDA terms were computed using up to 4 different functionals each, summing up to a total of 14 additional data points. These additional data were not included in any training or validation process but were used solely for testing of the ML models as a separate hold-out test set.

## 3.2 Density functional theory (DFT) calculations

For EDA calculations, DFT was used in combination with different functionals and long-range corrections (Table S1 for a full list). All geometry optimizations and EDA calculations were carried out using the AMS21 software package. For functionals where there was no DFT-D3 dispersion correction available in AMS21,[34] the standalone DFT-D3 package by Grimme and co-workers[35-37] was used. The starting geometries for the molecules **1-6** and **12-15** were obtained by a manually formed guess and subsequent pre-optimization using the UFF force-field in AMS. The starting structures for **10** and **11** were taken from the S22+ data set and the starting structures for the molecules **7-9** were obtained from reference 38. All final geometry optimizations were carried out with tight optimization parameters, *i.e.*, $10^{-9}$ a.u. for SCF energy, $10^{-6}$ a.u. for the total energy convergence, and $10^{-4}$ a.u./Å for the gradient convergence. For the geometry optimizations, we tested the impact of the functional on EDA by optimizing all structures with PBE[14] and a few molecules (**3**, **4**, **7**, **11**) with all other 28 method combinations. All calculations were performed using scalar relativistic approaches (ZORA)[39,40] and the TZ2P basis set.[41] The fragments, charges and spin polarizations employed for the EDA calculations can be found in Table S2 in the SI and chosen parameters were verified by comparing them to literature (see section S2.1 in the SI).[38,42]

The relative orbital and electrostatic contributions, $\Delta E_{rel,orb}$ and $\Delta E_{rel,elstat}$, respectively, have been calculated as a percentage of the sum of both (eq (4)), whilst the relative dispersive contribution, $\Delta E_{rel.disp}$, has been calculated as a percentage of the total interaction energy (eq (5)).

$$\Delta E_{rel,\mathrm{elstat}} = \frac{\Delta \mathrm{E}_{\mathrm{elstat}}}{\Delta \mathrm{E}_{\mathrm{elstat}} + \Delta \mathrm{E}_{\mathrm{orb}}} * 100 \qquad \text{eq (4)}$$

$$\Delta E_{rel,\mathrm{disp}} = \frac{\Delta \mathrm{E}_{int}(\mathrm{disp})}{\Delta \mathrm{E}_{int}(\mathrm{disp}) + \Delta \mathrm{E}_{int}(\mathrm{elec})} * 100 \qquad \text{eq (5)}$$

Note that $\Delta E_{rel,orb}$ is equal to 100-$\Delta E_{rel,elstat}$.

$$\Delta E_{rel,\mathrm{orb}} = \frac{\Delta \mathrm{E}_{\mathrm{orb}}}{\Delta \mathrm{E}_{\mathrm{elstat}} + \Delta \mathrm{E}_{\mathrm{orb}}} * 100 = 100 - \Delta E_{rel,elstat} \qquad \text{eq (6)}$$

To evaluate the differences between the used method combinations, the mean values (standard deviation ($\sigma$) (see SI section S2.2)) as well as the total energy ranges covered were calculated for each EDA term.

## 3.3 ML models

In this work, supervised and unsupervised ML to analyze and fit data obtained from EDA were used. For the input to ML models, we tested different combinations of EDA terms, PCs obtained from SOAP[32] applied to the chemical bond under investigation and a description of the functional number. PCA was needed to reduce the dimension of the SOAP descriptor as it contained over 16.000 values, while the descriptor based on EDA terms or the functional only contained 1-8 values. For the structural descriptor, 3 PCs were used that describe over 90% of the variance in the data (see Figure S2 in section S2.3 of the SI). The SOAP descriptor was generated using DScribe.[43] ML models were used as implemented in scikit-learn[43] using default parameters unless stated otherwise. The functional was encoded by a number between 1 and 28, which were subsequently normalized. For all features used to define the descriptor, the following normalization procedure was applied: $x' = \frac{x - min(x)}{max(x) - min(x)}$, with $x$ being the original feature space and $x'$ the standardized feature.

Hyperparameters for all ML models used were screened using 5-fold cross-validation.[29] For cross-validation, the data set was split into 75% training data, of which 90% were used for training and 10% for validation to obtain hyperparameters. 25% of the data were used as a hold-out test set to assess the performance of the ML models. As this 25% of data contained molecules with EDA terms obtained from different functionals, but structures already included in the training data, an additional test set with molecules never seen by the models was used to assess the performance.

For pattern recognition, clustering and classification models were tested. For clustering, k-means,[44] DBSCANk,[45] OPTICS,[46,47] Feature Agglomeration,[48] Spectral Clustering [49] and Affinity Propagation[50] were tested with minor deviations between models. Hence, k-means clustering was applied for all subsequent analyses. For classification tasks, we tested a multi-layer perceptron and support vector machine. The final study was carried out with the multi-layer perceptron. To investigate the influence of the functional on the bonding interpretation, we used a descriptor that encoded the EDA terms and the functional. Feature importance analysis[51,52] was used to assess the importance of each descriptor value for the bonding recognition task.

We further investigated the ability of ML to learn EDA terms obtained from different functionals and to apply these machine-learned values as inputs for classification. Therefore, supervised regression models were trained on EDA terms using a structural descriptor. As regressors, we tested kernel ridge regression, neural networks, and Gaussian process regression. Kernel ridge regression was found to give the lowest errors; hence these models were chosen for subsequent studies. All hyperparameters screened and final hyperparameters used are summarized in the SI in section S2.4.

# 4 Results and Discussion

## 4.1 EDA from different method combinations

To understand the impact of using different functionals on EDA results, we followed these steps: First, we averaged the individual energy terms of each molecule separately, considering all the functionals used. Second, we compared how these average values changed among molecules within the same bonding group. Finally, we compared these trends with the largest and smallest values obtained from the different functionals/methods tested for each energy term. This helps us to see if the trends remain consistent with varying method combinations and whether EDA terms of diverse studies and methods can be used to compare bonding patterns between molecules. Additionally, we calculated the mean, maximum, and minimum energy values for each molecule and energy type, which can be found in Tables S3-S13 in section S3.1 of the SI. EDA results when using the same functionals for the geometry optimization as in the EDA calculation are discussed and shown for a few examples in section S3.2 in the SI (Tables S14-S17). Effects can be considered negligible, justifying the use of PBE-D3 for geometry optimization of all systems.
When looking at the averaged energy terms of the bonding classes, one would assume that the bonding classes are well separated from each other. However, this is hardly the case and is visualized in Figure 3 using boxplots (bonding classes are indicated using the color scheme of Figure 2). In each panel of the figure, one box refers to one molecule. Each box shows where most energy terms obtained from the calculations, i.e., two quartiles, lie. The points indicate outliers. Outliers are mainly due to energy terms generated with the functionals OLYP, revPBE and M06-L but in general, energies of 16 method combinations were at least once regarded as outliers by the box-plot analysis (see Table S18 in section S3.3).

The plots can be interpreted for each EDA term individually. Take for instance the first panel that shows the distribution of $\Delta E_{int}$ of the different molecules. Covalently bound molecules (purple) show a considerable spread in the energy and are well separated from the rest of the classes (spreads can additionally be found in Table S19 in section S3.4 in the SI). In addition, hydrogen bonded systems (yellow) are well separated, but transition metal (green) and main group donor-acceptor bonded systems (blue) are overlapping. A similar result is obtained for $\Delta E_{orb.}$ With regards to $\Delta E_{elstat}$ and $\Delta E_{Pauli}$ covalently bound systems (purple), transition metal (green) and main group donor-acceptor (blue) bonds, are overlapping. Only hydrogen bound systems are well separated for all the four mentioned EDA terms. In case of $\Delta E_{int}$(disp), all bonding classes are overlapping.
The last panel, which shows results for $\Delta E_{rel,elstat}$, is a combination of $\Delta E_{orb}$ and $\Delta E_{elstat}$. As can be seen, combining these two values leads to an overall better separating of all the bonding classes observed, especially for the transition metal (green) and main group donor-acceptor (blue) groups, which are overlapping with other groups everywhere else. In addition, the covalently bound (purple) and main group donor-acceptor bound systems (blue) are well separated from each other. However, the hydrogen bound systems (yellow) are overlapping with the transition metal donor-acceptor group (green). The reason that these classes are overlapping is because the ratio between $\Delta E_{orb}$ and $\Delta E_{elstat}$ is approximately 1:3 for both classes (see e.g., Tables S9 to S13), hence the relative terms fall within the same range. Nevertheless, this effect does not limit the interpretability of the results as yellow classes are separated for all other terms except for $\Delta E_{int}$(disp).

In addition to the EDA conducted using average values, we also analyzed how the magnitudes of the EDA terms changed from one molecule to another (see Tables S20-S22). In summary, changes in energy from one molecule to another can be reproduced when looking only at results from methods that give minimum values or maximum values. Only in case of $\Delta E_{rel,elstat}$ an inverse trend can be observed when comparing maximum values of $H_3B\text{-}NH_3$ and $Cl_3B\text{-}NH_3$. An inverse trend means that instead of an energy decrease from one molecule to another, an energy increase or vice versa is observed. When comparing minimum/maximum values to each other instead of average values, the situation changes. For many molecules trends for $\Delta E_{rel,elstat}$, $\Delta E_{Pauli,}$ and $\Delta E_{int}$ change (the exact terms and molecules that are affected can be found in Table S22). In case of $\Delta E_{orb}$ and $\Delta E_{elstat}$, trends between only 2 systems, i.e., **7** to **9**, cannot be preserved. With respect to $\Delta E_{int}$(disp), no trend can be preserved. The same is true for $\Delta E_{rel,disp}$, as shown in the last panel of Figure 3. The latter might be caused by the fact that in most cases, a large part of the dispersion interaction is not captured by the functionals, but the external dispersion correction. Hence, for M06 the additional correction term is small, leading to $\Delta E_{int}$(disp) values close to zero, while the average dispersion energy for all molecules and all methods is -7.78 kJ/mol.
To summarize, when considering multiple EDA terms, such as $\Delta E_{int}$, $\Delta E_{orb}$, and $\Delta E_{elstat}$, or both $\Delta E_{int}$ and $\Delta E_{rel,elstat}$ of a bond, the four bonding classes can be separated even when looking at data obtained from different functionals. In general, $\Delta E_{int}$(disp) does not allow for separation of the bonding classes when comparing various methods.

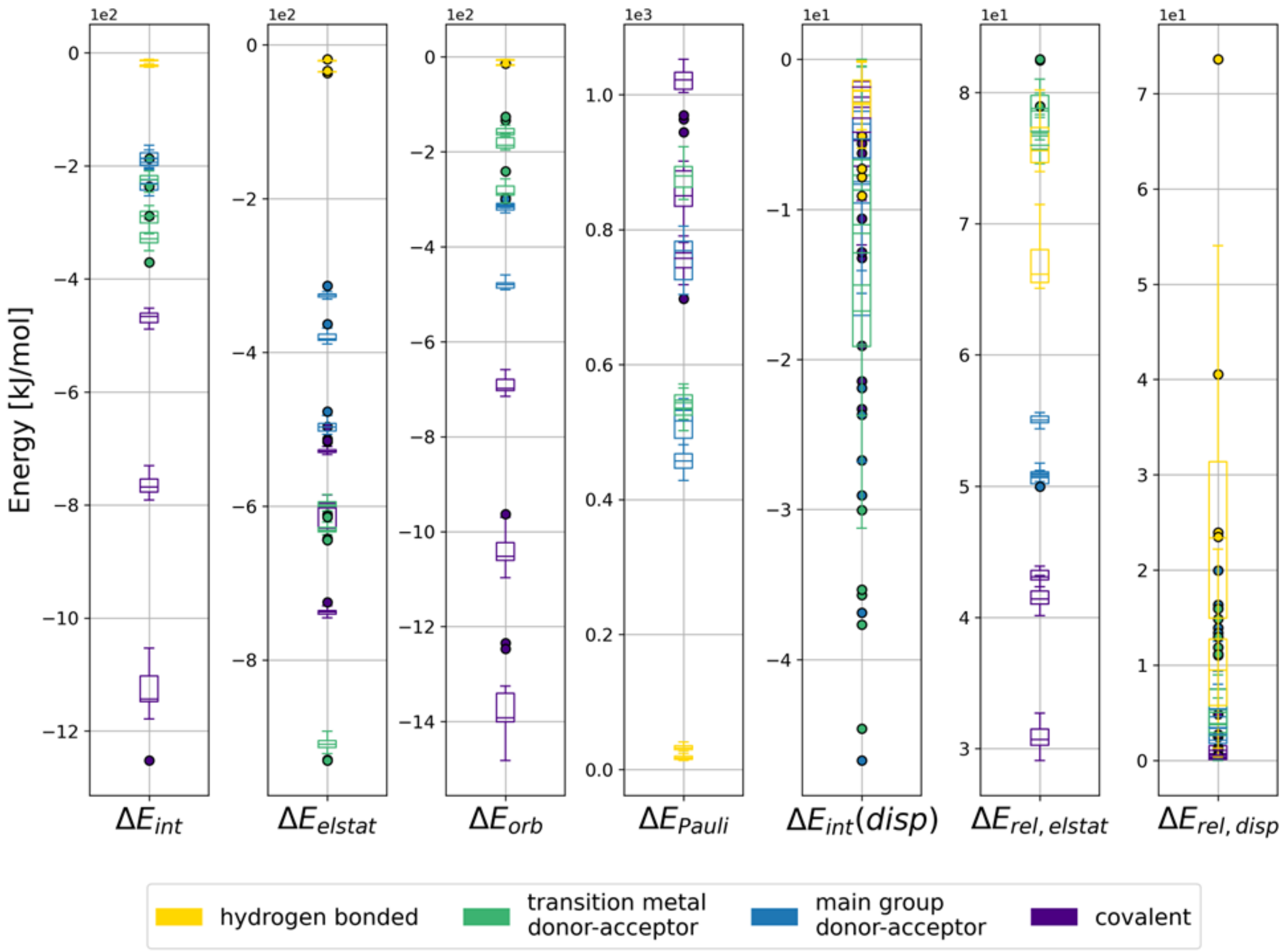

**Figure 3.** Boxplots of EDA energies of 28 distinct method-combinations for each molecule within the test set (see Figure 2). Molecules of the same bonding class are colored according to the legend in the bottom. For each energy term and molecule, the boxes show where two quartiles lie, while outliers are shown with dots.

## 4.2 Assessment of the influence of functionals via pattern recognition and feature importance analysis

As obtained from the previous analyses of EDA results, different functionals can influence the contributions of the EDA terms. Still, the previous analyses give unconcise results and do not allow to draw a clear conclusion about whether the amount of spread in the contributions leads to different bonding interpretations. Therefore, we applied pattern recognition that allows the grouping of the different molecules based on a predefined descriptor. Therefore, the descriptor encodes the different functionals in addition to properties that are important for the bond under investigation, i.e., the EDA terms. Details are discussed in the Methods section. In combination with feature importance analysis, this tool provides a way to investigate the influence of the functional on the pattern recognition into different bonding classes.

A multi-layer perceptron (MLPC) was used for the final analysis as it gave robust results for grouping the molecules with respect to their bonding types. To train the classifier, the training set was randomly split into 75% training data and 25% test data, meaning the model learned on 75% of the data how to classify molecules into four bonding groups. The trained model was then tested on the rest of the data with results shown in Figure 4. The colors in the plot refer to the different groups specified earlier, i.e., covalently bound molecules (purple), main group donor-acceptor bonds (blue), transition metal donor-acceptor groups (green), and hydrogen-bonded systems (yellow). All molecules in the test set were classified correctly, which can be seen by looking at the numbers and comparing to the original groups in Figure 2. As EDA terms for each molecule were computed 28 times, i.e., with 28 different methods, we had 28 different input vectors and thus data points for each molecule. Therefore, we computed an additional test set with completely new molecules (Figure 2, **12-15**). These molecules are also classified correctly and are marked with crosses in Figure 4.

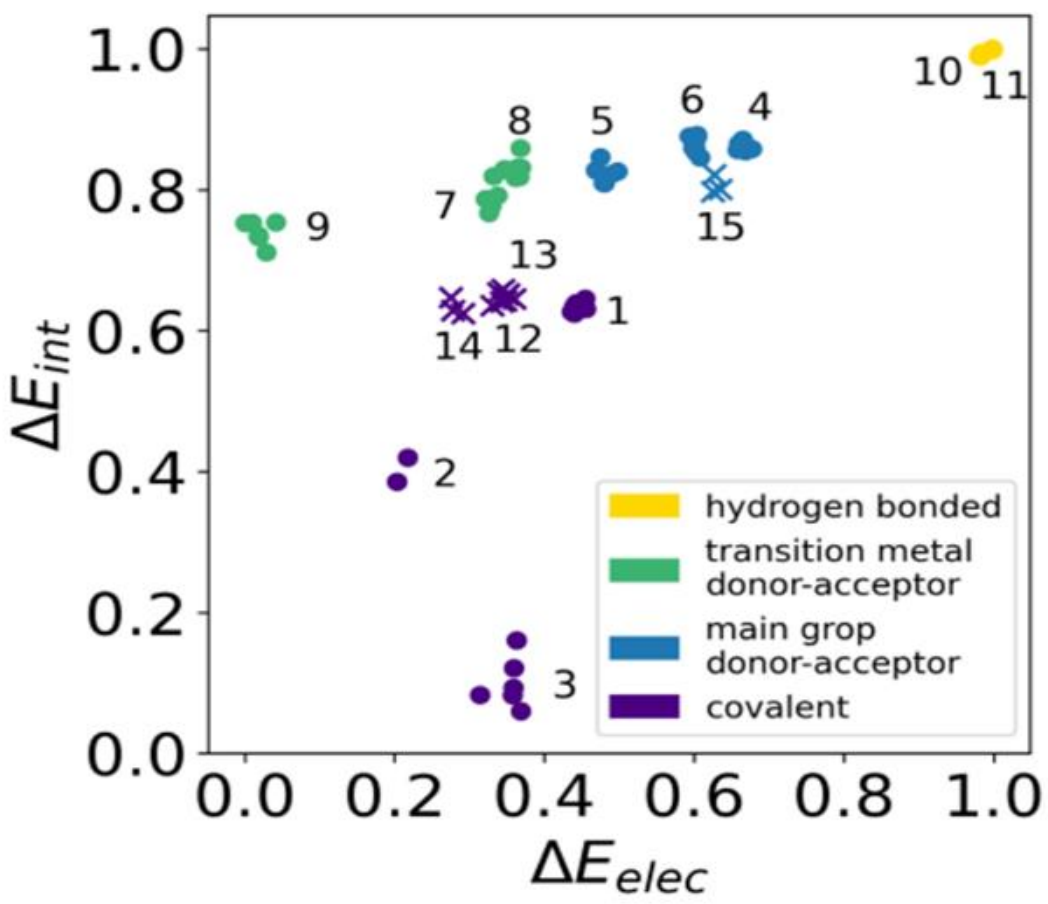

**Figure 4.** Normalized interaction energy plotted against normalized electrostatic energy for the molecules in test set 1 (25% of original datapoints) shown as dots and in test set 2 (completely new structures) shown as crosses. The color indicates the group as obtained by the trained MLPC model. Corresponding labels are shown in the legend. Numbers correspond to molecules in Figure 2.

As the model can successfully distinguish the different molecules based on EDA terms and functionals, we further sought to investigate the impact of each of the features on the pattern recognition task. Therefore, the permutational importance of each input feature was computed by observing how exchanging the values of a feature with random noise affects the error. If the error undergoes a significant change upon permutation, it indicates that the feature holds importance for our model. The procedure was carried out 100 times delivering a standard deviation value. The results are shown in Figure 5a. As can be seen, the largest contributions to the pattern recognition task come from $\Delta E_{rel,elstat}$ and $\Delta E_{Pauli}$. Interestingly, $\Delta E_{int}$(disp) and $\Delta E_{rel,disp}$ have very little influence, as can be seen from the small contributions in Figure 5a. This is encouraging as the previous discussion has shown that these terms do not allow for separation of the bonding classes. As can be further seen, the smallest contribution to the classification comes from the functional.

The influence of the functional was further analyzed by training the same classification algorithm on data of only one functional and applying it for classification of the rest of the data obtained from different functionals. Therefore, the EDA terms of the different functionals were used as an input and the functional was removed from the descriptor. This procedure should give rise to information about whether data of one functional can be used to correctly classify input data of other functionals or not. The results when encoding the functional in addition to the EDA terms led to similar trends and can be found in Figures S3 and S4. The results are illustrated in Figure 5b by using the percentage of incorrectly classified systems per trained algorithm. As can be seen, when training on EDA data obtained from most functionals, the errors for predicting molecules based on EDA terms obtained from other functionals are minor. However, there are two functionals that show over 20% error, i.e., OLYP-D3(BJ) and rPBE-D3(BJ), which have the largest error bars in Figure 5b. These results suggest that these functionals can lead to different bonding interpretations when comparing to other functionals. Among the functionals that could be used to train the MLPC model and classify almost all data of other functionals are PBE0, B3LYP, and variants thereof including different dispersion corrections, which have the smallest error bars in Figure 5b. These results are encouraging as indeed, these functionals are among the most widely used. In addition, Figure 5c shows the permutational importance of the descriptor values when training on one functional, which is shown using box plots.

As can be seen, the importance of each feature is comparable to the permutational importance when training on all functionals, as shown in Figure 5a. The main difference arises from a larger importance of $\Delta E_{orb}$ for the classifier trained on individual data of a single functional. These results further support that the functional has little influence on the bonding group classification.

Finally, we investigated for which molecules the bonding category was mostly predicted incorrectly by training MLPCs on data obtained from some of the worst (rPBE-D3 BJ, revPBE-D3 BJ, OLYP-D3 BJ, B97-D3 BJ) and best method combinations (PBE-D3 BJ and B3LYP-D3 BJ) individually (termed misclassification). The results are shown in Figure 5d. As can be seen, most molecules that are predicted incorrectly when using data from other method combinations as input to the MLPCs trained on one functional belong to groups of main group donor-acceptor bonds (blue). The functionals that lead to most misclassifications show up to over 80% of wrong predictions of this bonding type. A few transition metal donor-acceptor groups (green, abbreviated as TM D-A in the figure) were not classified correctly. No trend can be found for functionals that lead to relatively few errors. Hydrogen-bonded systems (yellow) are always classified correctly.
These results indicate that the main group and transition metal donor-acceptor groups might be better combined instead of separated, which is also supported by the EDA analysis in the previous section as these groups are overlapping for most EDA terms. From a chemical perspective, this makes sense as these groups are representing the same bonding type and are only distinguished by the type of atoms involved (resulting from either having d-orbitals in the valence space or not). Therefore, we repeated the whole analysis by choosing only three groups in total instead of four (see Figures S5 and S6 in section S4.2 in the SI). While analysis of the feature importance (Figure S5) leads to similar results with respect to the dispersion energy contributions and the functional importance, the importance of $\Delta E_{rel,elstat}$ decreased considerably when training on only 3 classes. This change can be explained by looking at $\Delta E_{rel,elstat}$ values of each of the molecules in the different groups in the last panel of Figure 3. It can be seen that $\Delta E_{rel,elstat}$ values of transition metal donor-acceptor molecules are approximately twice as large as the values of main group donor-acceptor molecules. When training a classification model on data of only one functional using three classes instead of four, the errors when predicting classes using data obtained from the other functionals are comparable to previous results using four classes (Figure S4 in the SI, section S4.2). As a large difference in $\Delta E_{rel,elstat}$ between transition metal and main group donor-acceptor bonds was found, four classes were determined more adequate for the conducted analysis.

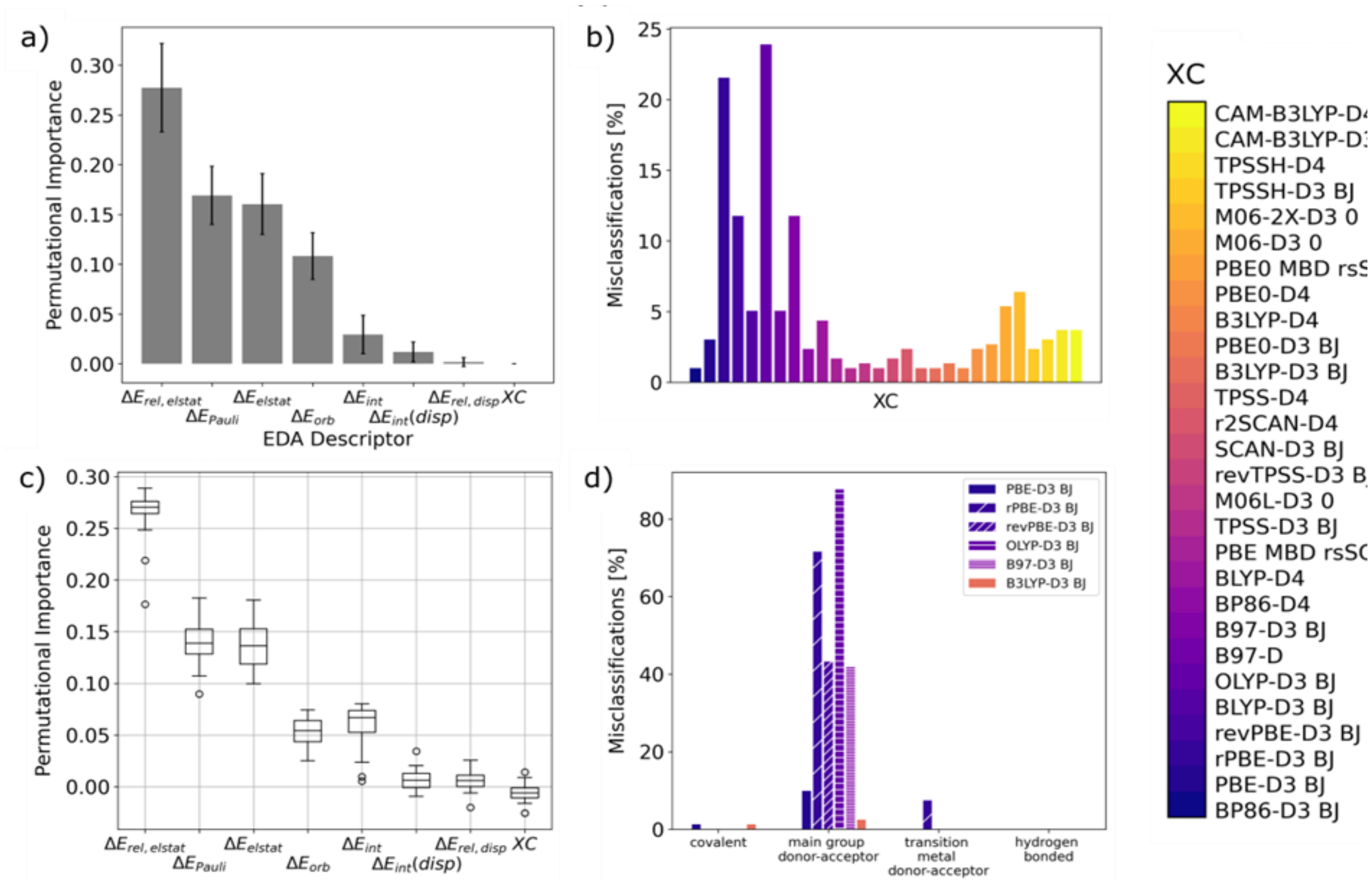

**Figure 5:** a) Permutational feature importance and standard deviation of each descriptor value for the classification task. b.) Percent of misclassifications for data when trained on data of one functional at a time. c) Feature importance for the MLPC trained on one functional at a time (functional trained on indicated by the color). d) Incorrectly predicted bond classes for a multi-layer perceptron (MLPC) separately trained on data of different functionals as indicated in the legend. Main group donor-acceptor and transition metal donor-acceptor bonds are abbreviated as MG D-A and TM D-A, respectively.

# 5 Summary and Outlook

This study investigates the impact of density functionals on the bonding interpretation through energy decomposition analysis (EDA). To address this, we generated a dataset representing four bonding groups comprising main group donor-acceptor bonds, transition metal donor-acceptor bonds, covalently bonded, and hydrogen-bonded systems. 28 calculations for 11 molecules using different functionals and long-range corrections were conducted. The comparison of density functional theory calculations revealed significant variation among the different functionals for various EDA terms, with the dispersion correction terms showing the most significant variation. Furthermore, the results show that the bonding classes are not always well separated from each other when looking at individual EDA terms, and there is some overlap between different bonding types, especially transition metal donor-acceptor and main group donor-acceptor bonds. However, relative energy terms that combine multiple EDA terms, help in better separating these bonding classes.

To answer the question of how much the functional choice influences the bonding analysis, a classification model was trained that takes EDA terms and the applied method as inputs to identify patterns and classify data into their respective bonding groups. This allowed examination of the importance of the features, i.e., the EDA terms and functional, for the classification task. Findings revealed that despite the considerable energy spread of the dispersion terms, they ranked second-to-last in contribution with the functional ranking last. This means the choice of the functional is not critical for the interpretation of the bonding analysis in terms of bonding classes.

Further analyses showed that data obtained using the most frequently applied functionals in EDA studies, especially PBE0, B3LYP, and their variations, are interchangeable when used as inputs for classification models, which is not the case when taking data of rPBE and OLYP functionals, for instance. These results indicate that selection of the latter functionals may not be optimal when applying EDA and comparing to studies using other methods.

## 5.1 Data availability:

All EDA calculations conducted in this study are uploaded to the open data repository NOMAD.[53] The EDA terms are further summarized in a csv file and are part of the supplementary information.

## 5.2 Code availability:

A jupyter notebook for EDA classification tasks can be found on github (https://github.com/ToOest/ML4EDA) including processed data for machine learning algorithms.

## 5.3 ORCID

Toni Oestereich https://orcid.org/0009-0006-4144-6457

Ralf Tonner-Zech https://orcid.org/0000-0002-6759-8559

Julia Westermayr https://orcid.org/0000-0002-6531-0742

## 5.4 Acknowledgment

We thank GOETHE-CSC Frankfurt, ZIH Dresden, PC2 Paderborn and HLR Stuttgart for computing time. We gratefully acknowledge the support of SCM/Amsterdam by providing a developer's license for the AMS code to RTZ.

# Supporting Information

Decoding Energy Decomposition Analysis: Machine-Learned Insights on the Impact of the Density Functional on the Bonding Analysis"

Toni Oestereich, Ralf Tonner-Zech*, Julia Westermayr*

Wilhelm-Ostwald-Institut für Physikalische und Theoretische Chemie, Universität Leipzig, 04103 Leipzig, Germany

corresponding authors: Prof. Dr. Julia Westermayr: Julia.westermayr@uni-leipzig.de; Prof. Dr. Ralf Tonner-Zech ralf.tonner@uni-leipzig.de

## Contents

# S1 Data set characteristics

As can be seen from Figure S1, the bond distances separate the different bonding groups well. The shortest bond belongs to the group of covalently bound systems, while the longest bonds are characteristic for hydrogen bonded molecules.

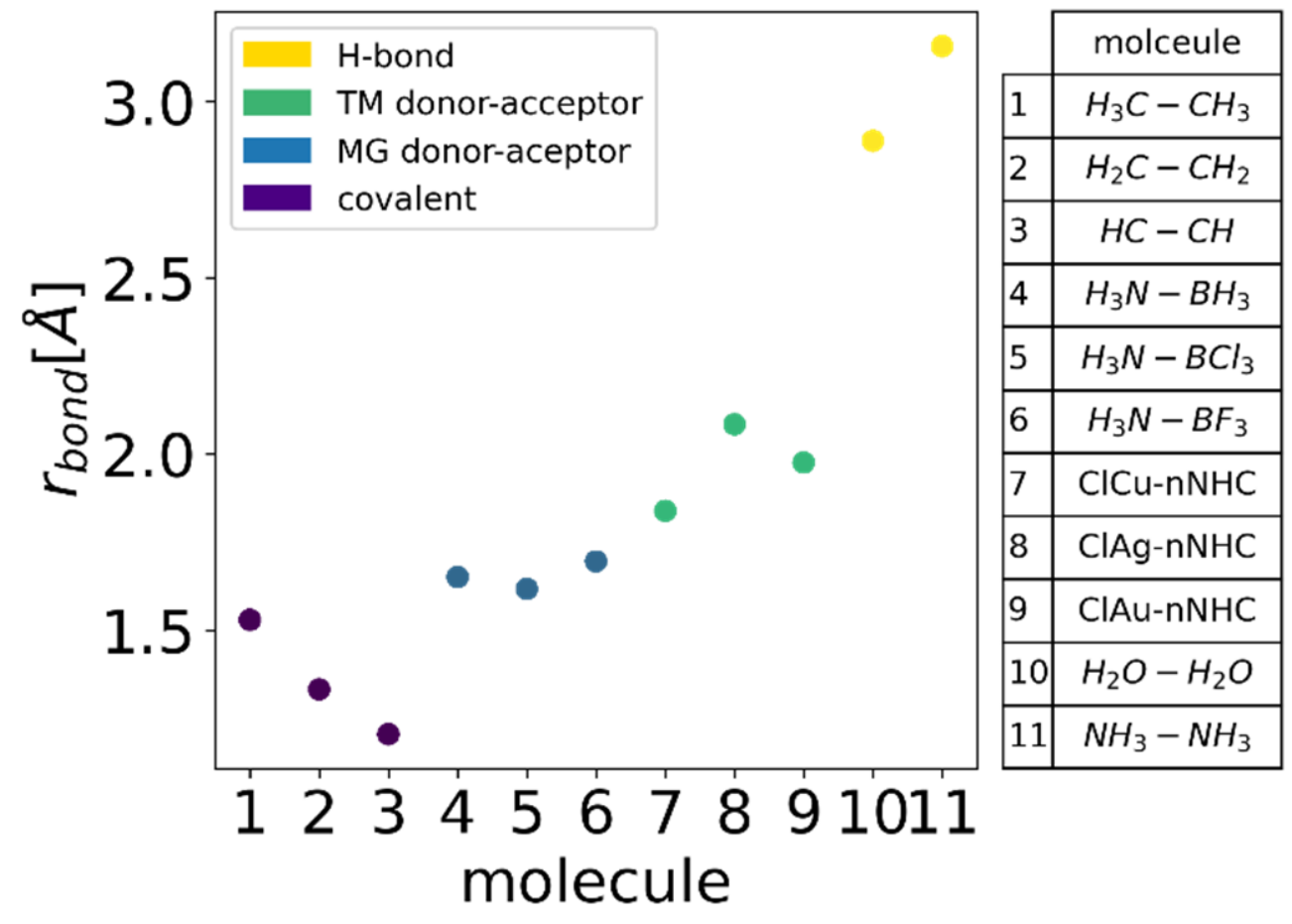

***Figure S 1**: Relation between the bond distance and the bonding group.*

# S2 Computational Details

## S2.1 Quantum chemical calculations

As explained in the Methods section of the main text, different methods were used to compute EDA terms. Therefore, 28 combinations of functionals and long-range corrections were applied. The scaling factors for the dispersion correction used in this work were compared to the ones given at the webpage of the university Bonn[1] ("List of Functionals and Coefficients for Zero-Damping – Chemie Universität Bonn" n.d., "List of Functionals and Coefficients for BJ-damping Chemie Universität Bonn" n.d.). For the functional revTPSS, the parameters of TPSS were used, as supported in the literature.[2] The functionals and additional parameters used are summarized in Table S1.

***Table S1**: Damping function and software packages used for the dispersion correction of a given functional.*

| Method no. | Functional | Dispersion correction | libXC[b] |
|---|---|---|---|
| 1 | BP86 | DFT-D3(BJ) | no |
| 2 | PBE | DFT-D3(BJ) | no |
| 3 | rPBE | DFT-D3(BJ)[a] | no |
| 4 | revPBE | DFT-D3(BJ) | no |
| 5 | BLYP | DFT-D3(BJ) | no |
| 6 | OLYP | DFT-D3(BJ)[a] | no |
| 7 | B97-D | | yes |
| 8 | B97 | DFT-D3(BJ)[a] | yes |
| 9 | BP86 | DFT-D4 (BJ) | no |
| 10 | BLYP | DFT-D4 | no |
| 11 | PBE | MBD rsSC | no |
| 12 | TPSS | DFT-D3(BJ) | no |
| 13 | M06L | DFT-D3(0) [a] | no |
| 14 | revTPSS | DFT-D3(BJ)[a] | no |
| 15 | SCAN | DFT-D3(BJ) | no |
| 16 | r2SCAN | DFT-D4 | yes |
| 17 | TPSS | DFT-D4 | no |
| 18 | B3LYP | DFT-D3(BJ) | no |
| 19 | PBE0 | DFT-D3(BJ)[a] | no |
| 20 | B3LYP | DFT-D4 | no |
| 21 | PBE0 | DFT-D4 | no |
| 22 | PBE0 | MBD rsSC | no |
| 23 | M06 | DFT-D3(0) [a] | no |
| 24 | M06-2X | DFT-D3(0) [a] | no |
| 25 | TPSSH | DFT-D3(BJ)[a] | no |
| 26 | TPSSH | DFT-D4 | no |
| 27 | CAM-B3LYP | DFT-D3(BJ) | yes |
| 28 | CAM-B3LYP | DFT-D4 | yes |

[a]The dispersion energy has been computed using the stand-alone version of DFT-D3 code (version 3.2 Rev 0, downloaded from Ref. [1]).

[b]These functionals have been computed using the libXC library as implementen in AMS.[3]

The fragments, charges and spin polarizations employed for the EDA calculations are summarized in Table S2.

***Table S2**: Fragmentation of a given molecule for the EDA calculations, charges, and spin polarization.*

| Molecule | Fragments | charge | spin polarization |
|---|---|---|---|
| Ethane | $CH_3/CH_3$ | 0/0 | +1/-1 |
| Ethene | $CH_2/CH_2$ | 0/0 | +2/-2 |
| Ethyne | CH/CH | 0/0 | +3/-3 |
| $H_3B$-$NH_3$ | $BH_3/NH_3$ | 0/0 | 0/0 |
| $Cl_3B$-$NH_3$ | $BCl_3/NH_3$ | 0/0 | 0/0 |
| $F_3B$-$NH_3$ | $BF_3/NH_3$ | 0/0 | 0/0 |
| ClCu-nNHC | CuCl/nNHC | 0/0 | 0/0 |
| ClAg-nNHC | AgCl/nNHC | 0/0 | 0/0 |
| ClAu-nNHC | AuCl/nNHC | 0/0 | 0/0 |
| Water dimer | $H_2O/H_2O$ | 0/0 | 0/0 |
| Ammonia dimer | $NH_3/NH_3$ | 0/0 | 0/0 |

## S2.2 Statistical analysis

To analyze the EDA results of 28 different functional and dispersion correction combinations for 11 molecules, the mean values, $\bar{E}$, maxima, minima, and standard deviations, $\sigma$, were calculated for the EDA results of each molecule. The mean and standard deviation of the different EDA energy terms were calculated according to the following formulas:

$$\Delta\bar{E} = \frac{1}{n}\left(\sum_{n=1}^{n} \Delta E_i\right) \qquad \text{eq (1)}$$

$$\sigma = \sqrt{\frac{1}{n}\sum_{n=1}^{n}(\Delta E_i - \Delta\bar{E})^2} \qquad \text{eq (2)}$$

## S2.3 Principal component analysis

In order to reduce the dimensionality of the SOAP descriptor,[4] containing over 1600 values per molecule, principal component analysis (PCA) was performed. The first three principal components (PCs) described 90 % of the variance of the SOAP descriptor.

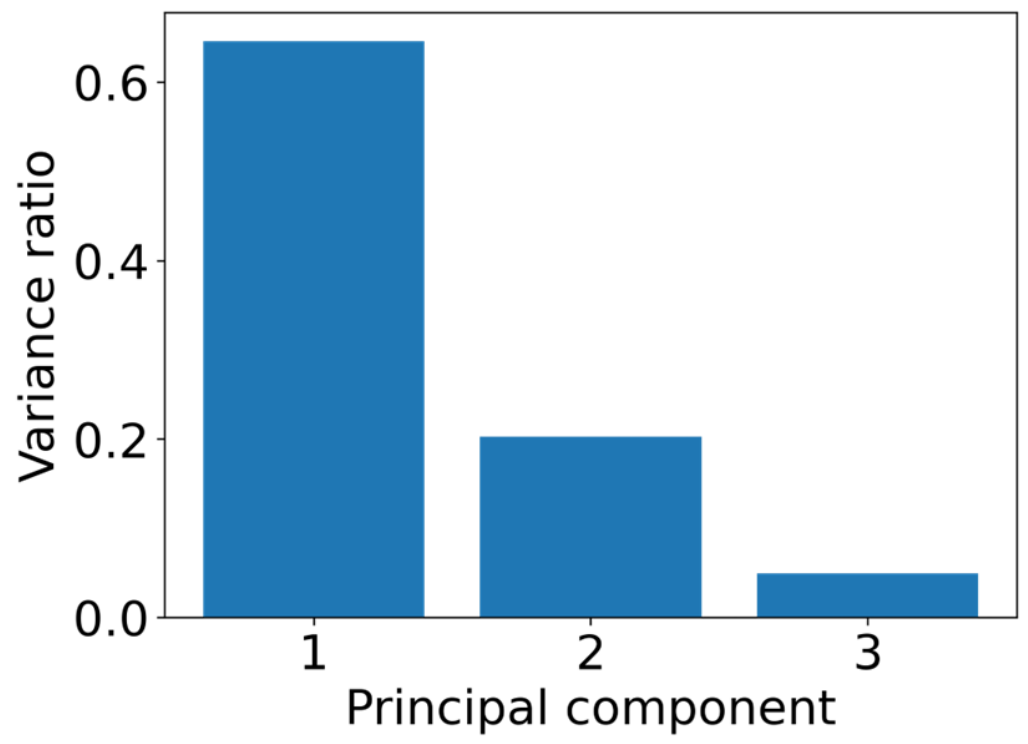

***Figure S 2**: Explained variance ratio of principle components 1 to 3 for the structural descriptor SOAP.[4]*

## S2.4 Model parameters

To choose suitable hyperparameters for the classification using a multi-layer perceptron for classification (MLPC), a grid search (regularization: 0.01- 0.000001, learning rate: 0.1-0.00000001, number of iterations: 100-7500) with 5-fold cross validation was performed on the trainingset, The input, i.e., descriptor, for the ML models were a combination of $\Delta E_{int}$(disp), $\Delta E_{elstat}$, $\Delta E_{int}$ ,$\Delta E_{orb}$, $\Delta E_{Pauli}$, $\Delta E_{rel,elstat}$, $\Delta E_{rel,disp}$, and XC-number. All values were normalized. The output was the bonding class label, which was predefined. The initial parameter grid was adjusted in case the optimized hyperparameter was close to the specified bounds. For the MLPC the final hyperparameters used were: regularization: 0.1, learning rate: 0.00001, maximum number of iterations: 7000.

The same procedure was carried out for the kernel ridge regression model, for which we used 3 principle components of the SOAP descriptor defined for the two atoms of the bond and the XC-number as input. The outputs were EDA terms as used for the input of classification. The following hyperparameter ranges were tested: regularization 0.1-0.000001, kernel width:0.1-0.0001, kernel: radial basis function, Laplacian, polynomial. The final hyperparameters used were a regularization rate of 0.000010, a kernel width of 0.005 in combination with a Laplacian kernel.

# S3 Summary of performed EDA calculations

## S3.1 Statistical results

In the following **Table S3** to **Table S13** the mean, $\Delta\bar{E}$, standard deviation, $\sigma$, largest, MAX$^c$, and smallest, MIN$^c$, values, as well as the spread (difference between largest and smallest value) for the EDA calculations of all 28 method combinations are given on a per-molecule basis.

The relative dispersive and electrostatic EDA terms are calculated according to equations 4 and 5 in the main part of the paper and are given in parentheses in the tables bellow. The largest and smallest values for the relative contributions are given for the whole data set and are thus not necessarily correspond to the non-relative EDA term (i.e. ,different method) given before the parentheses.

***Table S3**: Averages and spread of Energy decomposition analysis of ethane, values given in kJ/mol.*

| | $\Delta E_{int}$ | $\Delta E_{int}(disp)$ | | $\Delta E_{Pauli}$ | $\Delta E_{elstat}$ | | $\Delta E_{orb}$ | | $\Delta E_{prep}$ |
|---|---|---|---|---|---|---|---|---|---|
| $\Delta\bar{E}$ | -469 | -6 | (1.2 %) | 755 | -526 | (43.2 %) | -692 | (56.8 %) | 61 |
| $\sigma$ | 9 | 5 | (1.0 %) | 20 | 8 | (0.5 %) | 16 | (0.5 %) | 9 |
| MAX$^c$ | -452 | -0 | (4.9 %) | 791 | -497 | (43.9 %) | -659 | (57.7 %) | 81 |
| MIN$^c$ | -489 | -23 | (0.0 %) | 698 | -533 | (42.3 %) | -715 | (56.1 %) | 44 |
| Spread | 37 | 23 | (4.8 %) | 93 | 36 | (1.6 %) | 56 | (1.6 %) | 37 |

***Table S4**: Averages and spread of Energy decomposition analysis of ethene, values given in kJ/mol.*

| | $\Delta E_{int}$ | $\Delta E_{int}(disp)$ | | $\Delta E_{Pauli}$ | $\Delta E_{elstat}$ | | $\Delta E_{orb}$ | | $\Delta E_{prep}$ |
|---|---|---|---|---|---|---|---|---|---|
| $\Delta\bar{E}$ | -765 | -5 | (0.7 %) | 1018 | -737 | (41.5 %) | -1040 | (58.5 %) | -0 |
| $\sigma$ | 18 | 5 | (0.6 %) | 24 | 4 | (0.7 %) | 33 | (0.7 %) | 18 |
| MAX$^c$ | -730 | -0 | (2.7 %) | 1052 | -725 | (43.2 %) | -963 | (59.9 %) | 26 |
| MIN$^c$ | -791 | -21 | (0.0 %) | 944 | -745 | (40.1 %) | -1097 | (56.8 %) | -35 |
| Spread | 61 | 21 | (2.7 %) | 108 | 20 | (3.1 %) | 134 | (3.1 %) | 61 |

***Table S5**: Averages and spread of Energy decomposition analysis of ethyne, values given in kJ/mol.*

| | $\Delta E_{int}$ | $\Delta E_{int}(disp)$ | | $\Delta E_{Pauli}$ | $\Delta E_{elstat}$ | | $\Delta E_{orb}$ | | $\Delta E_{prep}$ |
|---|---|---|---|---|---|---|---|---|---|
| $\Delta\bar{E}$ | -1131 | -3 | (0.3 %) | 856 | -611 | (30.8 %) | -1373 | (69.2 %) | -18 |
| $\sigma$ | 38 | 3 | (0.3 %) | 33 | 20 | (0.8 %) | 52 | (0.8 %) | 38 |
| MAX$^c$ | -1053 | -0 | (1.2 %) | 901 | -585 | (32.7 %) | -1235 | (70.9 %) | 104 |
| MIN$^c$ | -1252 | -13 | (0.0 %) | 781 | -644 | (29.1 %) | -1482 | (67.3 %) | -95 |
| Spread | 199 | 13 | (1.2 %) | 120 | 59 | (3.6 %) | 247 | (3.6 %) | 199 |

***Table S6**: Averages and spread of Energy decomposition analysis of H3B-NH3, values given in kJ/mol.*

| | $\Delta E_{int}$ | $\Delta E_{int}(disp)$ | | $\Delta E_{Pauli}$ | $\Delta E_{elstat}$ | | $\Delta E_{orb}$ | | $\Delta E_{prep}$ |
|---|---|---|---|---|---|---|---|---|---|
| $\Delta\bar{E}$ | -192 | -7 | (3.5 %) | 458 | -325 | (50.6 %) | -317 | (49.4 %) | 44 |
| $\sigma$ | 9 | 6 | (3.2 %) | 14 | 3 | (0.5 %) | 7 | (0.5 %) | 9 |
| MAX$^c$ | -172 | -0 | (13.9 %) | 481 | -313 | (51.8 %) | -300 | (50.1 %) | 57 |
| MIN$^c$ | -205 | -27 | (0.0 %) | 428 | -331 | (49.9 %) | -328 | (48.2 %) | 24 |
| Spread | 33 | 27 | (13.9 %) | 53 | 18 | (1.9 %) | 28 | (1.9 %) | 33 |

***Table S7**: Averages and spread of Energy decomposition analysis of Cl3B-NH3, values given in kJ/mol.*

| | $\Delta E_{int}$ | $\Delta E_{int}(disp)$ | | $\Delta E_{Pauli}$ | $\Delta E_{elstat}$ | | $\Delta E_{orb}$ | | $\Delta E_{prep}$ |
|---|---|---|---|---|---|---|---|---|---|
| $\Delta\bar{E}$ | -232 | -14 | (6.0 %) | 759 | -497 | (50.8 %) | -480 | (49.2 %) | 129 |
| $\sigma$ | 14 | 11 | (4.7 %) | 29 | 7 | (0.3 %) | 7 | (0.3 %) | 14 |
| MAX$^c$ | -203 | -0 | (20.0 %) | 805 | -477 | (51.2 %) | -460 | (50.0 %) | 151 |
| MIN$^c$ | -253 | -47 | (0.0 %) | 704 | -507 | (50.0 %) | -491 | (48.8 %) | 100 |
| Spread | 50 | 47 | (19.9 %) | 101 | 30 | (1.2 %) | 31 | (1.2 %) | 50 |

***Table S8**: Averages and spread of Energy decomposition analysis of F3B-NH3, values given in kJ/mol.*

| | $\Delta E_{int}$ | $\Delta E_{int}(disp)$ | | $\Delta E_{Pauli}$ | $\Delta E_{elstat}$ | | $\Delta E_{orb}$ | | $\Delta E_{prep}$ |
|---|---|---|---|---|---|---|---|---|---|
| $\Delta\bar{E}$ | -186 | -8 | (4.2 %) | 513 | -381 | (55.0 %) | -311 | (45.0 %) | 95 |
| $\sigma$ | 13 | 7 | (3.8 %) | 24 | 6 | (0.3 %) | 5 | (0.3 %) | 13 |
| MAX$^c$ | -163 | -0 | (16.4 %) | 550 | -363 | (55.6 %) | -297 | (45.6 %) | 116 |
| MIN$^c$ | -207 | -29 | (0.1 %) | 468 | -389 | (54.4 %) | -318 | (44.4 %) | 72 |
| Spread | 44 | 29 | (16.3 %) | 82 | 26 | (1.3 %) | 21 | (1.3 %) | 44 |

***Table S9**: Averages and spread of Energy decomposition analysis of ClCu-nNHC, values given in kJ/mol.*

| | $\Delta E_{int}$ | $\Delta E_{int}(disp)$ | | $\Delta E_{Pauli}$ | $\Delta E_{elstat}$ | | $\Delta E_{orb}$ | | $\Delta E_{prep}$ |
|---|---|---|---|---|---|---|---|---|---|
| $\Delta\bar{E}$ | -288 | -12 | (4.1 %) | 536 | -631 | (77.6 %) | -182 | (22.4 %) | -10 |
| $\sigma$ | 16 | 8 | (2.8 %) | 18 | 8 | (1.4 %) | 14 | (1.4 %) | 16 |
| MAX$^c$ | -236 | -0 | (11.9 %) | 565 | -611 | (82.5 %) | -133 | (23.6 %) | 22 |
| MIN$^c$ | -320 | -36 | (0.2 %) | 502 | -644 | (76.4 %) | -196 | (17.5 %) | -62 |
| Spread | 84 | 35 | (11.7 %) | 63 | 33 | (6.1 %) | 62 | (6.1 %) | 84 |

***Table S10**: Averages and spread of Energy decomposition analysis of ClAg-nNHC, values given in kJ/mol.*

| | $\Delta E_{int}$ | $\Delta E_{int}(disp)$ | | $\Delta E_{Pauli}$ | $\Delta E_{elstat}$ | | $\Delta E_{orb}$ | | $\Delta E_{prep}$ |
|---|---|---|---|---|---|---|---|---|---|
| $\Delta\bar{E}$ | -224 | -12 | (5.4 %) | 545 | -599 | (79.2 %) | -158 | (20.8 %) | -9 |
| $\sigma$ | 12 | 9 | (3.9 %) | 14 | 8 | (1.0 %) | 9 | (1.0 %) | 12 |
| MAX$^c$ | -187 | 0 | (15.9 %) | 571 | -585 | (82.5 %) | -126 | (21.9 %) | 14 |
| MIN$^c$ | -247 | -38 | (-0.0 %) | 519 | -614 | (78.1 %) | -168 | (17.5 %) | -46 |
| Spread | 60 | 38 | (15.9 %) | 53 | 29 | (4.4 %) | 42 | (4.4 %) | 60 |

***Table S11**: Averages and spread of Energy decomposition analysis of ClAu-nNHC, values given in kJ/mol.*

| | $\Delta E_{int}$ | $\Delta E_{int}(disp)$ | | $\Delta E_{Pauli}$ | $\Delta E_{elstat}$ | | $\Delta E_{orb}$ | | $\Delta E_{prep}$ |
|---|---|---|---|---|---|---|---|---|---|
| $\Delta\bar{E}$ | -328 | -14 | (4.3 %) | 879 | -910 | (76.4 %) | -282 | (23.6 %) | -6 |
| $\sigma$ | 16 | 10 | (3.0 %) | 20 | 11 | (1.0 %) | 15 | (1.0 %) | 16 |
| MAX$^c$ | -289 | -0 | (13.0 %) | 923 | -893 | (78.9 %) | -241 | (25.4 %) | 37 |
| MIN$^c$ | -370 | -45 | (0.2 %) | 845 | -931 | (74.6 %) | -308 | (21.1 %) | -45 |
| Spread | 81 | 44 | (12.8 %) | 78 | 38 | (4.4 %) | 68 | (4.4 %) | 81 |

***Table S12**: Averages and spread of Energy decomposition analysis of H2O dimer, values given in kJ/mol.*

| | $\Delta E_{int}$ | $\Delta E_{int}(disp)$ | | $\Delta E_{Pauli}$ | $\Delta E_{elstat}$ | | $\Delta E_{orb}$ | | $\Delta E_{prep}$ |
|---|---|---|---|---|---|---|---|---|---|
| $\Delta\bar{E}$ | -22 | -2 | (10.5 %) | 33 | -35 | (66.8 %) | -18 | (33.2 %) | -2 |
| $\sigma$ | 2 | 2 | (8.7 %) | 3 | 1 | (1.6 %) | 1 | (1.6 %) | 2 |
| MAX$^c$ | -19 | -0 | (40.6 %) | 41 | -34 | (71.5 %) | -14 | (34.9 %) | 1 |
| MIN$^c$ | -25 | -8 | (0.4 %) | 28 | -37 | (65.1 %) | -19 | (28.5 %) | -5 |
| Spread | 6 | 8 | (40.2 %) | 13 | 3 | (6.4 %) | 5 | (6.4 %) | 6 |

***Table S13**: Averages and spread of Energy decomposition analysis of NH3 dimer, values given in kJ/mol.*

| | $\Delta E_{int}$ | $\Delta E_{int}(disp)$ | | $\Delta E_{Pauli}$ | $\Delta E_{elstat}$ | | $\Delta E_{orb}$ | | $\Delta E_{prep}$ |
|---|---|---|---|---|---|---|---|---|---|
| $\Delta\bar{E}$ | -13 | -3 | (24.5 %) | 17 | -21 | (76.1 %) | -7 | (23.9 %) | -2 |
| $\sigma$ | 1 | 2 | (16.5 %) | 3 | 1 | (1.8 %) | 1 | (1.8 %) | 1 |
| $MAX^c$ | -12 | -0 | (73.6 %) | 24 | -19 | (80.2 %) | -5 | (26.0 %) | 0 |
| $MIN^c$ | -15 | -9 | (1.3 %) | 14 | -22 | (74.0 %) | -8 | (19.8 %) | -3 |
| Spread | 3 | 9 | (72.3 %) | 11 | 3 | (6.2 %) | 2 | (6.2 %) | 3 |

## S3.2 Geometry dependence on EDA calculations

To evaluate the influence of functional choice for the geometry optimization on the EDA results, the geometry of the molecules ethyne, ammonia borane, ClCu-nNHC and the ammonia dimer have been optimized with all 28 functional dispersion correction combinations, see Table S1. The subsequent EDA has been carried out with the same functional as used in the geometry optimization. Comparing the spreads (between different functionals) of the different energy terms for the PBE-D3 optimized calculations and those optimized with the same functional as used in the EDA the influence of the geometry optimization is evaluated. For the interaction, orbital, electrostatic and relative electrostatic energy no clear trends between the spreads have been found. Only for the dispersion energy spread of the calculations where the geometry was optimized with the same functional as used in the subsequent EDA calculation, the spread was either the same or lower relative to the one for the PBE-D3 optimized EDA (see Tables S3-S13 above).

***Table S14**: Averages and spread of Energy decomposition analysis of ethyne, optimized with same functionals as used in EDA, values given in kJ/mol.*

| | $\Delta E_{int}$ | $\Delta E_{int}(disp)$ | | $\Delta E_{Pauli}$ | $\Delta E_{elstat}$ | | $\Delta E_{orb}$ | | $\Delta E_{prep}$ |
|---|---|---|---|---|---|---|---|---|---|
| $\Delta\bar{E}$ | -1125 | -3 | (0.3 %) | 865 | -612 | (30.8 %) | -1375 | (69.2 %) | 3 |
| $\sigma$ | 32 | 3 | (0.3 %) | 24 | 16 | (0.9 %) | 47 | (0.9 %) | 1 |
| $MAX^c$ | -1055 | -0 | (1.2 %) | 907 | -591 | (33.1 %) | -1258 | (70.9 %) | 4 |
| $MIN^c$ | -1179 | -13 | (0.0 %) | 814 | -648 | (29.1 %) | -1450 | (66.9 %) | 1 |
| Spread | 124 | 13 | (1.2 %) | 94 | 57 | (4.0 %) | 192 | (4.0 %) | 3 |

***Table S15**: Averages and spread of Energy decomposition analysis of $H_3B$-$NH_3$, optimized with same functional as used in EDA, values given in kJ/mol.*

| | $\Delta E_{int}$ | $\Delta E_{int}(disp)$ | | $\Delta E_{Pauli}$ | $\Delta E_{elstat}$ | | $\Delta E_{orb}$ | | $\Delta E_{prep}$ |
|---|---|---|---|---|---|---|---|---|---|
| $\Delta\bar{E}$ | -190 | -7 | (3.6 %) | 449 | -320 | (50.7 %) | -312 | (49.3 %) | 66 |
| $\sigma$ | 10 | 6 | (3.2 %) | 12 | 10 | (0.6 %) | 11 | (0.6 %) | 29 |
| $MAX^c$ | -167 | -0 | (14.0 %) | 465 | -296 | (51.9 %) | -285 | (50.1 %) | 179 |
| $MIN^c$ | -203 | -27 | (0.0 %) | 424 | -333 | (49.9 %) | -329 | (48.1 %) | 50 |
| Spread | 35 | 27 | (14.0 %) | 41 | 38 | (2.0 %) | 44 | (2.0 %) | 129 |

***Table S16:*** *Averages and spread of Energy decomposition analysis of ClCu-nNHC, optimized with same functional as used in EDA, values given in kJ/mol.*

| | $\Delta E_{int}$ | $\Delta E_{int}(disp)$ | | $\Delta E_{Pauli}$ | $\Delta E_{elstat}$ | | $\Delta E_{orb}$ | | $\Delta E_{prep}$ |
|---|---|---|---|---|---|---|---|---|---|
| $\Delta\bar{E}$ | -289 | -12 | (4.0 %) | 510 | -611 | (77.5 %) | -177 | (22.5 %) | 2 |
| $\sigma$ | 14 | 8 | (2.7 %) | 37 | 29 | (1.4 %) | 19 | (1.4 %) | 3 |
| MAX$^{c}$ | -247 | -0 | (11.6 %) | 542 | -484 | (81.4 %) | -111 | (23.8 %) | 4 |
| MIN$^{c}$ | -320 | -35 | (0.2 %) | 348 | -654 | (76.2 %) | -196 | (18.6 %) | -11 |
| Spread | 73 | 34 | (11.4 %) | 194 | 170 | (5.2 %) | 85 | (5.2 %) | 14 |

***Table S17****: Averages and spread of Energy decomposition analysis of NH3 dimer, optimized with same functional as used in EDA, values given in kJ/mol.*

| | $\Delta E_{int}$ | $\Delta E_{int}(disp)$ | | $\Delta E_{Pauli}$ | $\Delta E_{elstat}$ | | $\Delta E_{orb}$ | | $\Delta E_{prep}$ |
|---|---|---|---|---|---|---|---|---|---|
| $\Delta\bar{E}$ | -13 | -3 | (23.4 %) | 16 | -20 | (76.6 %) | -6 | (23.4 %) | 0 |
| $\sigma$ | 1 | 2 | (12.8 %) | 3 | 3 | (1.9 %) | 1 | (1.9 %) | 3 |
| MAX$^{c}$ | -11 | -0 | (51.2 %) | 21 | -10 | (80.2 %) | -2 | (25.9 %) | 11 |
| MIN$^{c}$ | -15 | -7 | (1.3 %) | 6 | -23 | (74.1 %) | -8 | (19.8 %) | -7 |
| Spread | 4 | 7 | (49.9 %) | 15 | 13 | (6.1 %) | 5 | (6.1 %) | 18 |

## S3.3 Details for Figure 3 of the main text

***Table S18**: Functionals considered outliers in Figure 3 given for each EDA term. The molecule for which the EDA term of a given functional was considered an outlier is indicated in rounded brackets.*

| $\Delta E_{Pauli}$ | $E_{int}$ | $\Delta E_{elstat}$ | $\Delta E_{orb}$ | $\Delta E_{int}(disp)$ | $E_{rel,elstat}$ | $E_{rel,disp}$ |
|---|---|---|---|---|---|---|
| SCAN-D3 BJ (Ethene) | TPSS-D4 (Ethyne) | SCAN-D3 BJ (Ethane) | - | revPBE-D3 BJ (Ethane) | - | revPBE-D3 BJ (Ethane) |
| r2SCAN-D4 (Ethene) | SCAN-D3 BJ (ClAu-nNHC) | r2SCAN-D4 (Ethane) | - | B97-D3 BJ (Ethane) | - | B97-D3 BJ (Ethane) |
| BP86-D3 BJ (Ethyne) | - | r2SCAN-D4 (Ethene) | - | revPBE-D3 BJ (Ethene) | - | revPBE-D3 BJ (Ethene) |
| BLYP-D3 BJ (Ethyne) | - | BP86-D3 BJ (Ethyne) | - | B97-D3 BJ (Ethene) | - | B97-D3 BJ (Ethene) |
| B97-D3 BJ (Ethyne) | - | BLYP-D3 BJ (Ethyne) | - | revPBE-D3 BJ (Ethyne) | - | revPBE-D3 BJ (Ethyne) |
| SCAN-D3 BJ (Ethyne) | - | B97-D3 BJ (Ethyne) | - | B97-D3 BJ (Ethyne) | - | B97-D3 BJ (Ethyne) |
| r2SCAN-D4 (Ethyne) | - | TPSS-D4 (Ethyne) | - | BLYP-D4 (Ethyne) | - | BLYP-D4 (Ethyne) |
| BLYP-D3 BJ (ClCu-nNHC) | - | CAM-B3LYP-D3 BJ ($H_2O$- $H_2O$) | - | revPBE-D3 BJ ($NH_3$-$BH_3$) | - | revPBE-D3 BJ ($NH_3$-$BH_3$) |
| BLYP-D4 (ClCu-nNHC) | - | CAM-B3LYP-D4 ($H_2O$- $H_2O$) | - | B97-D3 BJ ($NH_3$-$BH_3$) | - | B97-D3 BJ ($NH_3$-$BH_3$) |
| SCAN-D3 BJ (ClCu-nNHC) | - | - | - | revPBE-D3 BJ ($NH_3$-$BF_3$) | - | revPBE-D3 BJ ($NH_3$-$BF_3$) |
| r2SCAN-D4 (ClCu-nNHC) | - | - | - | B97-D3 BJ ($NH_3$-$BF_3$) | - | B97-D3 BJ ($NH_3$-$BF_3$) |
| BLYP-D3 BJ (ClAg-nNHC) | - | - | - | revPBE-D3 BJ ($H_2O$- $H_2O$) | - | revPBE-D3 BJ ($H_2O$- $H_2O$) |
| BLYP-D4 (ClAg-nNHC) | - | - | - | B97-D3 BJ ($H_2O$- $H_2O$) | - | B97-D3 BJ ($H_2O$- $H_2O$) |
| SCAN-D3 BJ (ClAg-nNHC) | - | - | - | B97-D3 BJ ($NH_3$-$NH_3$) | - | B97-D3 BJ ($NH_3$-$NH_3$) |
| BLYP-D3 BJ (ClAu-nNHC) | - | - | - | - | - | - |
| BLYP-D4 (ClAu-nNHC) | - | - | - | - | - | - |

## S3.5 Relative spread of EDA terms

In order to evaluate which EDA term shows the largest difference between different calculation parameters the spread for each EDA term, relative to the average value for each molecule is given below.

***Table S19**: Relative spread in % of different EDA terms for each molecule. Average over all molecules given in the last row.*

| **Species** | $\Delta E_{int}$ | $\Delta E_{int}(disp)$ | $\Delta E_{Pauli}$ | $\Delta E_{elstat}$ | $\Delta E_{orb}$ |
|---|---|---|---|---|---|
| Ethane | 8 | 413 | 12 | 7 | 8 |
| Ethene | 8 | 417 | 11 | 3 | 13 |
| Ethyne | 18 | 439 | 14 | 10 | 10 |
| $H_3B$-$NH_3$ | 17 | 400 | 12 | 6 | 9 |
| $Cl_3B$-$NH_3$ | 22 | 338 | 13 | 6 | 6 |
| $F_3B$-$NH_3$ | 24 | 372 | 16 | 7 | 7 |
| ClCu-nNHC | 29 | 305 | 12 | 5 | 35 |
| ClAg-nNHC | 27 | 316 | 10 | 5 | 27 |
| ClAu-nNHC | 25 | 321 | 9 | 4 | 24 |
| $H_2O$ dimer | 27 | 345 | 40 | 9 | 28 |
| $NH_3$ dimer | 23 | 280 | 57 | 14 | 46 |
| **Average** | 21 | 359 | 19 | 7 | 20 |

## S3.5 Trends of EDA influenced by the choice of method

To assess whether the various methods can influence energy trends (as defined in the main part and below) between two molecules, we analyzed whether a trend changes or not when comparing two molecules in the following Tables. To obtain reference trend between two molecules we use the EDA terms for all calculation methods averaged. These trends are then compared to the trends either using only the calculation method giving the highest or lowest value for a given EDA term. If the sign of the energy difference between two molecule changes in comparison to the energy difference seen for the average over all calculation methods it will be interpreted as a misrepresented trend (indicated as "-"). If a trend is preserved, an "x" is shown in the Table. The results are summarized in the Tables S15 and S16.

***Table S20**: Comparison of trends (between two molecules) between the average of all calculation methods and for calculation method giving the highest value for each EDA term. Both trends having the same sign is indicated as a cross ("x"), whilst both trends showing different signs is indicated as minus ("-").*

| $E_{int}$ | $\Delta E_{elstat}$ | $E_{orb}$ | $\Delta E_{Pauli}$ | $\Delta E_{int}(disp)$ | $E_{rel,disp}$ | $E_{rel,elstat}$ | molecules |
|---|---|---|---|---|---|---|---|
| X | X | X | X | X | X | X | Ethane -> Ethene |
| X | X | X | X | X | X | X | Ethene -> Ethyne |
| X | X | X | X | X | X | X | Ethyne -> $NH_3BH_3$ |
| X | X | X | X | X | X | - | $NH_3BH_3$ -> $NH_3BCl_3$ |
| X | X | X | X | X | X | X | $NH_3BCl_3$ -> $NH_3BF_3$ |
| X | X | X | X | X | X | X | $NH_3BF_3$ -> ClCu-nNHC |
| X | X | X | X | X | X | X | ClCu-nNHC -> ClAg-nNHC |
| X | X | X | X | X | X | X | ClAg-nNHC -> ClAu-nNHC |
| X | X | X | X | X | X | X | ClAu-nNHC -> $H_2O$-$H_2O$ |
| X | X | X | X | X | X | X | $H_2O$-$H_2O$ -> $NH_3$-$NH_3$ |

***Table S21**: Comparison of trends (between two molecules) between the average of all calculation methods and for calculation method giving the lowest value for each EDA term. Both trends having the same sign is indicated as a cross ("x"), whilst both trends showing different signs is indicated as minus ("-").*

| $E_{int}$ | $\Delta E_{elstat}$ | $E_{orb}$ | $\Delta E_{Pauli}$ | $\Delta E_{int}(disp)$ | $E_{rel,disp}$ | $E_{rel,elstat}$ | molecules |
|---|---|---|---|---|---|---|---|
| X | X | X | X | X | X | X | Ethane -> Ethene |
| X | X | X | X | X | X | X | Ethene -> Ethyne |
| X | X | X | X | X | X | X | Ethyne -> $NH_3BH_3$ |
| X | X | X | X | X | X | X | $NH_3BH_3$ -> $NH_3BCl_3$ |
| X | X | X | X | X | X | X | $NH_3BCl_3$ -> $NH_3BF_3$ |
| X | X | X | X | X | X | X | $NH_3BF_3$ -> ClCu-nNHC |
| X | X | X | X | X | X | X | ClCu-nNHC -> ClAg-nNHC |
| X | X | X | X | X | X | X | ClAg-nNHC -> ClAu-nNHC |
| X | X | X | X | X | X | X | ClAu-nNHC -> $H_2O$-$H_2O$ |
| X | X | X | X | X | X | X | $H_2O$-$H_2O$ -> $NH_3$-$NH_3$ |

In the following the "worst case" scenario of comparing trends in EDA energies between molecules is assessed. To do so the average trends in EDA energies between two molecules are compared with the trend obtained when the EDA of one molecules has been performed with the calculation method giving the highest value for a EDA term and the EDA of the second molecule has been performed with the calculation method giving the lowest value for a EDA term. It is again considered a change in trend if the sign of the two trends is different.

**Table S22**: Comparison of trends (between two molecules) between the average of all calculation methods and for the calculation method giving the lowest value for one molecule and the calculation method giving the highest value for the other molecule. Both trends having the same sign is indicated as a cross (x), whilst both trends showing different signs is indicated as minus (-).

| $E_{int}$ | $\Delta E_{elstat}$ | $E_{orb}$ | $\Delta E_{Pauli}$ | $\Delta E_{int}(disp)$ | $E_{rel,disp}$ | $E_{rel,elstat}$ | **molecules** |
|---|---|---|---|---|---|---|---|
| X | X | X | X | - | - | - | Ethane -> Ethene |
| X | X | X | X | - | - | X | Ethene -> Ethyne |
| X | X | X | X | - | - | X | Ethyne -> $NH_3BH_3$ |
| - | X | X | X | - | - | - | $NH_3BH_3$ -> $NH_3BCl_3$ |
| - | X | X | X | - | - | X | $NH_3BCl_3$ -> $NH_3BF_3$ |
| X | X | X | - | - | - | X | $NH_3BF_3$ -> ClCu-nNHC |
| - | - | - | - | - | - | - | ClCu-nNHC -> ClAg-nNHC |
| X | X | X | X | - | - | - | ClAg-nNHC -> ClAu-nNHC |
| X | X | X | X | - | - | X | ClAu-nNHC -> $H_2O$-$H_2O$ |
| X | X | X | X | - | - | X | $H_2O$-$H_2O$ -> $NH_3$-$NH_3$ |

Notably the cooper and silver complexes seem to be the only cases where every EDA term can show a changing trend. Thus EDA trends between two molecules should be obtained by comparing EDA results of the same, or at least similar, calculation method.

# S4 Machine learning based assessment of functional importance

To evaluate the EDA terms that influence the classification the most, the permutational importance for the MLPC has been calculated employing the ELI5 package.[5] The permutational importance feature in ELI5 calculates the importance of each feature in an MLPC by permutation of the feature values with random noise and measuring the resulting change in the model's performance. The feature importance scores are then normalized so that they add up to 1.

The standard deviation of the feature importance scores is a measure of how much the importance score varies when the permutations are applied repeatedly. It is calculated by repeating the permutation process multiple times and calculating the standard deviation of the resulting feature importance scores. For the below calculations the number of iterations was set to 100 in order to ensure a meaningful permutational importance.

To ensure that the obtained feature importance is suitable for generalization it has been calculated for the test set (see Figure 2a of the main text).

## S4.1 MLPC results for 4 bonding classes excluding the XC number from the descriptor

The feature importance has been determined for an MLPC containing labels for 4 bonding classes with (Figure 5a) and without (Figure S5) the functional as a descriptor value.

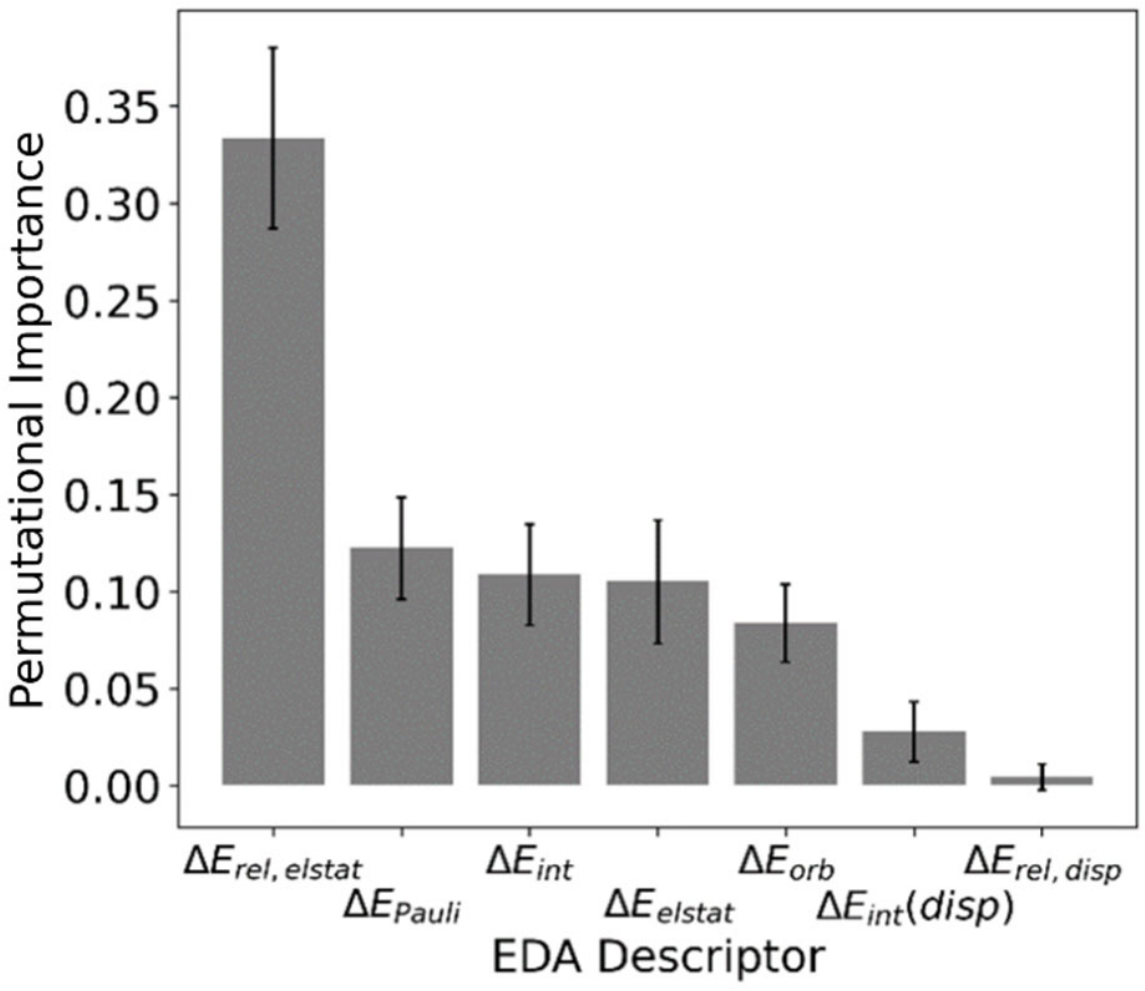

***Figure S 3**: Feature importance for MLPC trained on 4 bonding classes, excluding the XC number from the descriptor.*

To evaluate which of the 28 functional-dispersion-correction combinations might lead to wrong classifications of the bond type the MLPC has been trained on each of the combinations individually and was then tested on the remaining 27 combinations. Figure S4 (in addition to Figure 5b in the main text) shows the number of misclassifications of the training set labeled with 4 bond classes.

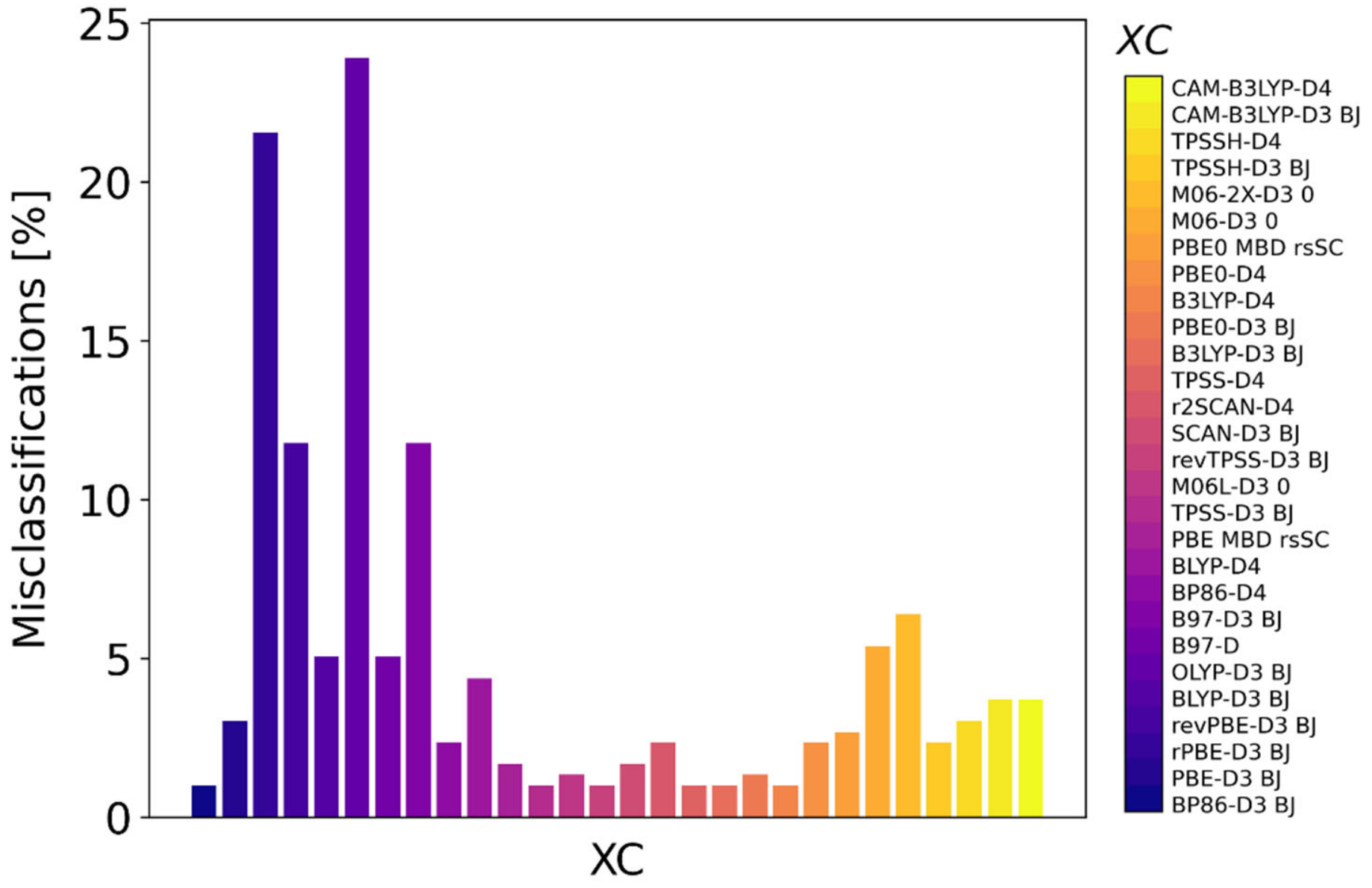

***Figure S 4**: Number of wrong classifications for 4 bonding classes, excluding the XC number from the descriptor.*

## S4.2 MLPC results for 3 bonding classes

In addition, we performed the previous analyses using only 3 bonding classes (Figure S5 and S6). The descriptor includes information on the method chosen.

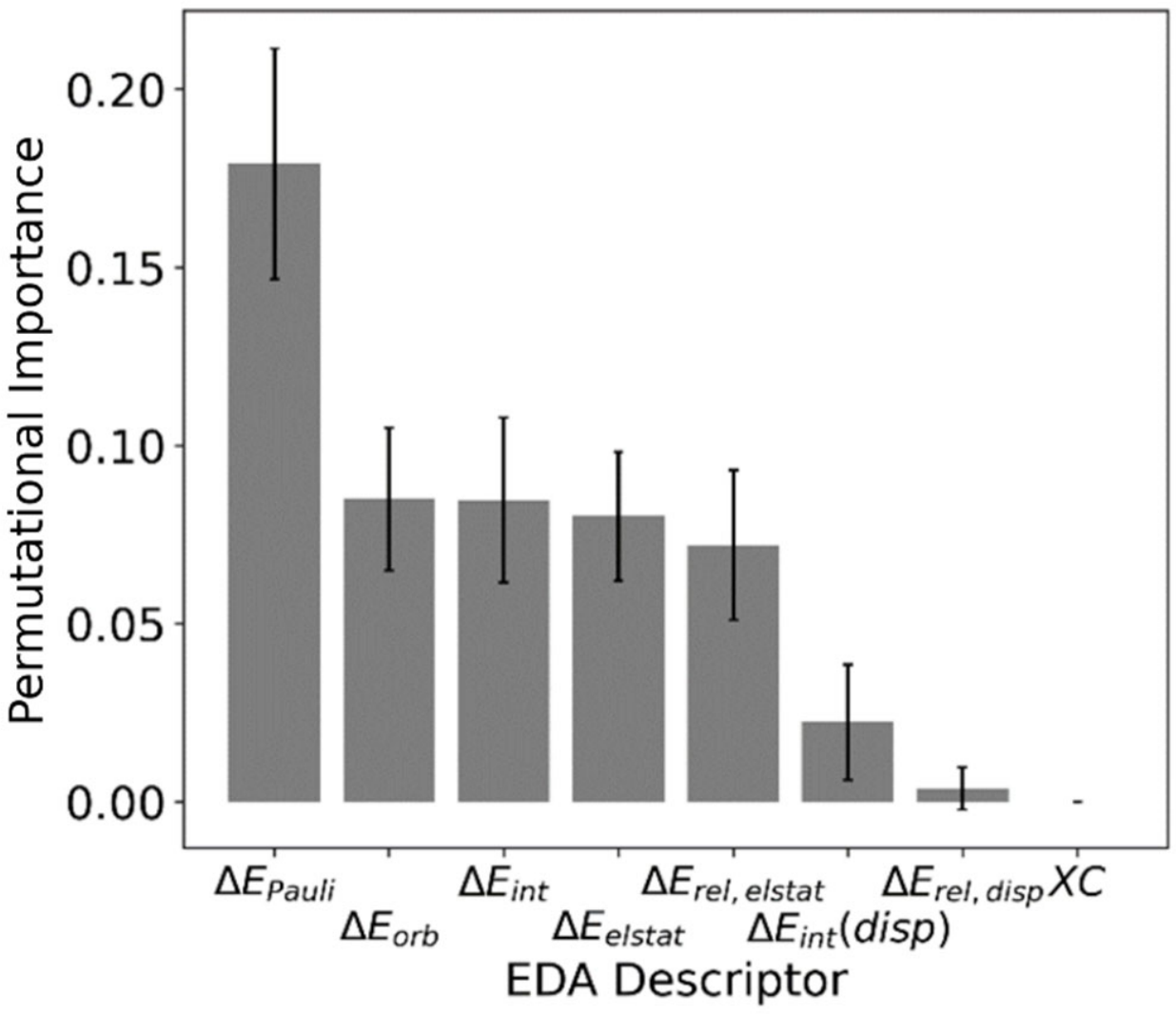

***Figure S 5***: *Feature importance for MLPC trained on 3 bonding classes with the XC number included in the descriptor.*

Whilst for 4 bond classes the relative electrostatic energy seems to have the biggest impact on the classification, the Pauli energy has the biggest influence for 3 bond classes. For both classifications the XC number as well as well as the relative dispersion energy show a negligible permutational importance.

As it is visible from Figure S6, the number of misclassifications hardly changes when using 3 bonding classes instead of 4 (compare to Figure 5b in the main text).

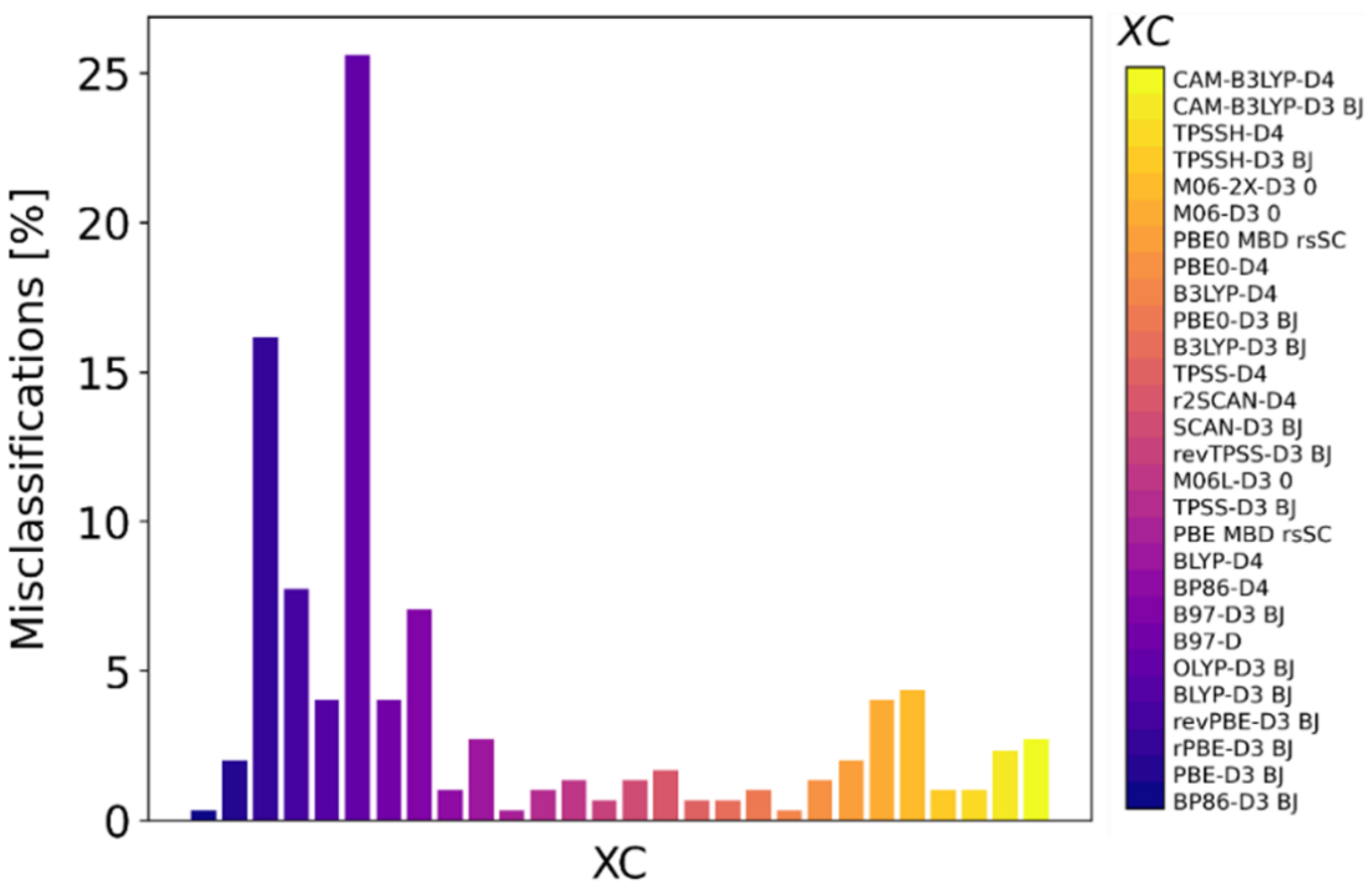

***Figure S 6***: *Number of wrong classifications, with functional number for 3 Bonding classes.*